\documentclass[11pt]{scrartcl}

\usepackage{preamble_W}

\newcommand{\setOfReals}{\mathbb{R}}
\newcommand{\setOfNaturals}{\mathbb{N}}
\newcommand{\setOfNonnegativeIntegers}{{\mathbb{N}_0}}
\newcommand{\setOfPositiveReals}{{\setOfReals_{+}}}

\newcommand{\borel}[1]{\mathcal{B} (#1 )}

\newcommand{\sys}[1]{\textsc{#1}}

\newcommand{\measureIntegral}[2]{\langle#1, #2 \rangle }
\newcommand{\measureIntegralSum}[2]{\langle \langle#1, #2 \rangle \rangle}

\makeatletter
\let\oldabs\abs
\def\abs{\@ifstar{\oldabs}{\oldabs*}}

\newcommand{\cadlag}{c\`adl\`ag\,}

\newcommand{\defeq}{:=}

\newcommand{\indicator}[1]{\mathsf{1}_{\{#1\}}}

\newcommand{\distribution}[2]{\sys{#1}\left(#2\right)}

\newcommand{\differential}[1]{\,\mathrm{d} #1}
\newcommand{\timeDerivative}[1]{\frac{\differential}{\differential t} #1 }
\newcommand{\partialDerivative}[2]{\frac{\partial}{\partial #1} #2 }

\newcommand{\eqstop}{.}
\newcommand{\eqcomma}{,}

\newcommand{\E}{\mathsf{E}}
\newcommand{\Eof}[1]{\E\left[#1 \right]}

\newcommand{\prob}{\mathsf{P}}
\newcommand{\probOf}[1]{\prob\left(#1\right)}

\newcommand{\myExp}[1]{\exp \left( #1 \right)  }

\newcommand{\ie}{\textit{i.e.}}
\newcommand{\eg}{\textit{e.g.}}

\renewcommand {\theequation}{\arabic{section}.\arabic{equation}}

\title{Dynamic Survival Analysis for non-Markovian Epidemic Models}

\author{Francesco Di Lauro$^{1, \star}$, Wasiur R. KhudaBukhsh$^{2, \star}$\hspace{1.5mm}\href{https://orcid.org/0000-0003-1803-0470}{\includegraphics[width=3mm]{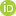}},  Istv{\'a}n Z. Kiss$^3$,\\ Eben Kenah$^4$\hspace{1.5mm}\href{https://orcid.org/0000-0002-7117-7773}{\includegraphics[width=3mm]{ORCID-iD_icon-16x16}}, Max Jensen$^3$\hspace{1.5mm}\href{https://orcid.org/0000-0002-9973-472X}{\includegraphics[width=3mm]{ORCID-iD_icon-16x16}}, Grzegorz A. Rempa{\l}a$^4$\hspace{1.5mm}\href{https://orcid.org/0000-0002-6307-4555}{\includegraphics[width=3mm]{ORCID-iD_icon-16x16}}\\
\\
$^1$\sys{University of Oxford}, $^2$\sys{University of Nottingham},\\ $^3$\sys{University of Sussex}, $^4$\sys{The Ohio State University}\\ 
$^\star${\small Both authors contributed equally and are joint first authors.}}

\date{}

\begin{document}
    \maketitle
    
    \begin{abstract}
        We present a new method  for analyzing stochastic epidemic models under minimal assumptions.  The method, dubbed \ac{DSA},  is based on a simple yet powerful observation, namely  that  population-level mean-field trajectories described by a system of \acp{PDE}  may also approximate  individual-level  times of infection and recovery. This idea  gives rise to a certain non-Markovian agent-based model and  provides  an agent-level likelihood function for a random sample of infection and/or recovery times. Extensive numerical analyses on both synthetic and real epidemic data from the \ac{FMD} in the United Kingdom and the COVID-19 in India show good accuracy and confirm method's versatility in likelihood-based parameter estimation. The   accompanying software package gives prospective users a  practical tool for  modeling,  analyzing   and interpreting epidemic data with the help of the DSA approach.  
    \end{abstract}
    
    
    \section{Introduction}
    The standard approach to building a stochastic compartmental epidemic model is to make use of \acp{CTMC} to keep track of the sizes of the compartments over time (\eg, number of individuals with different immunological statuses) using counting processes (see \cite{anderson_britton}). Following the random time change representation of Poisson processes \cite{ethier2009markov,anderson_kurtz_CRN}, the trajectory equations for those counting processes are written in terms of independent, unit rate Poisson processes. When the size of the population under consideration is large, those counting processes, appropriately scaled, converge to deterministic, continuous real-valued functions satisfying certain \acp{ODE} by virtue of the \ac{FLLN} for Poisson processes (see \cite{kurtz_1970_solutions,Kurtz_1978_Strong}). This provides a link between the stochastic and the deterministic world. Those limiting \acp{ODE} are often referred to as the mean-field equations in the literature. Famous examples include the classical Kermack--McKendrick equations for the \ac{SIR} epidemic model \cite{Kermack1927Contribution}.   
    
    However, this  astounding popularity of the standard Markov models or the corresponding mean-field \ac{ODE} models seems to belie their apparent lack of faithfulness to the underlying biology of the disease. Indeed, the population count-based Markov models assume exponentially distributed inter-event times. As a consequence, the instantaneous rates of infection and recovery are assumed constant regardless of key epidemiologically relevant covariates, such as the age of infection, time since vaccination etc. While there are more advanced stochastic models that do incorporate those covariates (as we will also  do  in this paper), those models are often fit to data in an ad hoc fashion; or are too computationally expensive to be useful for practical purposes. Our aim  in this work is to build a principled and rigorous statistical approach to fitting those more advanced stochastic models to data without compromising on simplicity. 
    
    In this paper, we present a survival analytic approach,  dubbed  \acf{DSA}, that constructs probability distributions of individual times of infection and recovery from population-level (mean-field) trajectory equations. In \cite{KhudaBukhsh2019SDS}, a subset of the authors first employed this idea in the context of the classical Kermack--McKendrick Markovian \ac{SIR} epidemics described by their mean-field \acp{ODE}. Here, we extend the idea to the vastly more realistic class of non-Markovian models that allow non-exponential contact interval \cite{kenah2011biostat} and infectious periods. The theoretical underpinning is laid down by an extension of the so-called \emph{Sellke construction} \cite{Sellke1983asymptotic,anderson_britton}, which we describe in detail in \Cref{sec:sellke}. 
    
    There are several advantages of \ac{DSA}. First, \ac{DSA} does not require knowledge of the size of the susceptible population, which is almost always unknown in real epidemics and often assumed to be the population of the entire city, state, or even a country. In fact, \ac{DSA} not only avoids this ad hoc adjustment, but also provides a ready estimate of the \emph{effective population size}, tracking of which could provide further insights into an ongoing epidemic. Second, \ac{DSA} does not require the whole epidemic trajectory and works with only a random sample of infection and, if available, recovery times. Third, on the strength of its survival analytic foundation, \ac{DSA} is able to handle censoring, truncation and aggregation of data (over time and population) in a straightforward manner. 
    
    The rest of the paper is structured as follows: \Cref{sec:model} describes the stochastic model in terms of measure-valued processes along with their large population mean-field limits. In \cref{sec:dsa}, we describe the Sellke construction and the \ac{DSA} approach in detail before conducting extensive numerical analysis in \Cref{sec:numerical}. We apply the \ac{DSA} method to the \acf{FMD} in the United Kingdom and the COVID-19 in India. In \Cref{sec:numerical}, we also provide synthetic data analysis  so that \ac{DSA} could be compared against ground truth. Finally, we conclude with a short discussion in \Cref{sec:discussion}. For the sake of completeness, additional mathematical derivations and numerical figures are provided in the Appendix. We will adhere to the following conventions about notations and symbols throughout the paper.  
    

    
    
    \paragraph{Notational conventions} We denote the sets of natural numbers, non-negative integers, real numbers and non-negative real numbers by $\setOfNaturals, \setOfNonnegativeIntegers, \setOfReals$,  and $\setOfPositiveReals$ respectively. The set of Borel subsets of a set $E$ will be denoted by $\borel{E}$.  For a set $E$, we use the notation $D([0,\infty), E)$ (or $ D([0,T], E)$) to denote the space of $E$-valued \cadlag functions defined on $[0,\infty)$ (or $[0,T]$, for some $T >0$). The stochastic processes that we consider in this paper will be elements of $D([0,\infty), E)$ or $ D([0,T], E))$ for some state space $E$ and some time horizon $T>0$ unless otherwise specified. The set-function $\delta_{x}$ is the Dirac measure, \ie, for a set $A$, the function $\delta_{x}(A)$ takes value $1$ if $x \in A$ and $0$ otherwise.  For a point measure $\nu=\sum_{i=1}^n\delta_{x_i}$ and a measurable function $f$, the integration of the function $f$ with respect to the measure $\nu$ will be denoted by
    \begin{align*}
        \langle\nu,f\rangle \defeq \int f\,d\nu = \sum_{i=1}^n f(x_i).
    \end{align*}
    For a vector of point measures $\nu \defeq (\nu_1, \nu_2, \ldots, \nu_k)$, for some positive integer $k$, and a measurable function $f$, we use the notation $\measureIntegralSum{\mu}{f}$ to denote 
    \begin{align*}
       \measureIntegralSum{\nu}{f} \defeq \sum_{i=1}^{k} \measureIntegral{\nu_i}{f} \eqstop 
     \end{align*}
     The indicator (or characteristic) function of a set $A$ is denoted by $\indicator{A}$, \ie, $\indicator{A}(x) = 1$ if $x\in A$ and $0$ otherwise.  
     Other notations will be introduced when required.

    \section{Stochastic model}
    \label{sec:model}
    Because we want to keep track of important epidemiological covariates along with counts of individuals in different compartments, our primary tool will be measure-valued processes, which are naturally capable of carrying more information than raw population counts. The measure-valued representation will also allow us to turn an inherently non-Markovian model into a Markov model, albeit on a more abstract state space. While the age of infection is the most natural choice for ``age'', one may also use the notion of age to account for other important covariates that describe time since some specific event. For instance, the biological age, time since vaccination are important for certain infectious diseases. Therefore, we use the term ``age'' in a broad sense and keep track of the ages of individuals with different immunological statuses (susceptible, infected, recovered/removed). 
    
    \paragraph{Measure-valued processes} Suppose we have $n$ susceptible and $m$ infected individuals initially. We assume $m$ depends on $n$ in the sense that $m/n \rightarrow \rho$ as $n \rightarrow \infty$ for some $\rho  \in (0,1)$.  Let us now define the following stochastic processes
    \begin{align}
    \begin{aligned}
    X_{t}^{S} & \defeq \sum_{k=1}^{N_{S}(t)} \delta_{s_{k}(t)} \eqcomma \quad 
    X_{t}^{I} & \defeq \sum_{k=1}^{N_{I}(t)} \delta_{i_{k}(t)} \eqcomma \quad 
    X_{t}^{R} & \defeq \sum_{k=1}^{N_{R}(t)} \delta_{r_{k}(t)} \eqcomma 
    \end{aligned}
    \label{eq:measure_valued_processes}
    \end{align}
    where $N_S(t), N_{I}(t)$, and $ N_{R}(t)$ are  the total numbers of susceptible,  infected, and recovered individuals in the population at time $t$. The quantities $s_k(t), i_k(t)$, and $ r_k(t)$ are the ages of the $k$-th susceptible, infected, and recovered  individual (following some specific ordering convention).    The measure-valued stochastic processes $X_{t}^{S}, X_{t}^{I}$, and $ X_{t}^{R}$ keep track of the age distribution of the population of individuals. For instance, taking the ``age'' for the infected individuals to represent the age of infection,  $X_{t}^{I}(A)$ gives us the number of infected individuals whose ages of infection lie in the set $A$. To be precise, the processes $X_{t}^{S}, X_{t}^{I}$, and $ X_{t}^{R}$ are finite, point-measures on $\setOfPositiveReals$ with atoms placed on the individual ages. Therefore, we have the following self-consistency relations $N_{S}(t) = \measureIntegral{X_{t}^{S}}{1} = X_{t}^{S} \left(\setOfPositiveReals\right)$, $N_{I}(t) = \measureIntegral{X_{t}^{I}}{1} = X_{t}^{I} \left(\setOfPositiveReals\right)$, and $N_{R}(t) = \measureIntegral{X_{t}^{R}}{1} = X_{t}^{R}\left(\setOfPositiveReals\right)$, where $1$ is the identity function. Now, define the stochastic process
    \begin{align}
        X_t \defeq (X_{t}^{S}, X_{t}^{I}, X_{t}^{R}) \eqcomma 
    \end{align} 
    which describes the dynamics of the infectious disease at the population level. We also have the conservation law: $\measureIntegralSum{X_t}{1} = n+m$. The process $X_t$ is a Markov process with paths in $D([0,T], \mathcal{M}_{P}(\setOfPositiveReals)^3 )$ where $T>0$ is a finite time horizon and $\mathcal{M}_{P}(\setOfPositiveReals)$ is the space of finite, point measures on $\setOfPositiveReals$.
    Although we do not explicitly show the dependence of the stochastic process $X_t$ on the initial size of the susceptible population $n$, it is worth keeping in mind. 
    
    We adopt the pairwise model of \cite{kenah2011biostat} to describe the dynamics of the epidemic process under the stochastic mass-action set-up. There are two types of events: Infection and natural recovery. In order to describe the intensities (of the Markov process $X_t$) corresponding to these two types of events, let us introduce two functions: $\beta: \setOfPositiveReals\times\, \setOfPositiveReals \to \setOfPositiveReals$ and $\gamma: \setOfPositiveReals \to \setOfPositiveReals$. The function $\beta(u,v)$ describes the instantaneous intensity of an infectious contact between a susceptible individual of age $u$ and an infectious individual of age $v$. That is, the probability that a susceptible individual of age $u$ will be infected by an infectious individual of age $v$ in the next $\delta t$ time unit  is $n^{-1}\beta(u, v) \delta t$ under the stochastic law of mass-action, where $\delta t$ is assumed infinitesimally small.  For each $u \in \setOfPositiveReals$, we shall often treat $\beta(u, \bullet)$ and $\beta(\bullet, u)$ as real functions. In the language of the pairwise model \cite{kenah2011biostat}, the function $\beta$ characterizes the probability law of the \emph{contact intervals}. The function $\gamma$ is the hazard function that characterizes the probability law of the infectious period. Note that neither of these two probability laws needs to be exponential, even though $X_t$ itself is a Markov process (see \cite{KhudaBukhsh2020Delay} for a similar example in the context of a \ac{CRN}). The infection and natural recovery processes are assumed independent.  We also assume recovered individuals can no longer infect others or be infected.  
    
    From the classical theory of stochastic epidemiology, we know that appropriately scaled population counts in \ac{CTMC}-based epidemic models converge to solutions to \acp{ODE} in the large population (mean-field) limit (see \cite{anderson_britton}). They are a consequence of the \ac{FLLN}-type approximation theorems for Markov processes \cite{kurtz_1970_solutions,Kurtz_1978_Strong}. The intuition is that the stochastic fluctuation, which is typically described in terms of a zero-mean martingale after a Doob--Meyer decomposition of the counting processes around the mean vanishes in the limit. A similar intuition holds true for measure-valued Markov processes. Indeed, the scaled process $n^{-1}X_t$ converges to a vector of deterministic measure-valued functions in the limit of $n \rightarrow \infty$. Furthermore, when the limiting measure-valued functions admit densities, it is possible to describe them using \acfp{PDE}. We describe the limiting system in the following. 

    \subsection{Mean-field limit}
    We are interested in the limit of the epidemic process as $n \to \infty$ with $n/m \to \rho$, for some $\rho \in (0,1)$. Therefore, in the limit, the total scaled population size is $(1+\rho)$. We scale the system this way because we wish to interpret the susceptible curve as a survival function, which takes the value one at zero. We shall make this point more elaborate in \cref{sec:dsa} on \ac{DSA}. 
    
    Under some technical assumptions on the intensities and the initial population size (more precise statement in \cref{appendix:derivation}),  the scaled stochastic process $n^{-1}X_t$ converges to a vector-valued deterministic continuous function $x_t \defeq (x_t^S, x_t^I, x_t^R)$, where the components $x_t^S, x_t^I$, and $ x_t^R$ are measure-valued functions. A brief, intuitive sketch of the proof of convergence of the scaled process $n^{-1}X_t$ to the deterministic function $x_t$ is provided in \cref{appendix:derivation} for the sake of completeness. The main technical tools are borrowed from   existing probability theory literature on Banach space-valued Markov processes. In particular, similar techniques and derivations can be found in \cite{Fournier2004microscopic,Champagnat2008individual,Tran2008limit,Tran2009traits,Meleard:2012:SFS}. While the limiting measure-valued functions can be evaluated against a large class of test functions whence various moments can be calculated, they are not necessarily easy to work with from a practical perspective.  

    When the limiting system of measure-valued deterministic functions $x_t^{S}, x_t^{I}$, and $x_t^{R}$ admit densities $y_S(t, \bullet), y_I(t, \bullet)$, and $y_R(t, \bullet)$  with respect to the Lebesgue measure, we can describe the densities in terms of the following system of \acp{PDE}:
    \begin{align}
    \begin{aligned}
        \left(\partial_t + \partial_s \right)y_{S}(t,s) &{} = - y_{S}(t,s) \int_0^\infty \beta(s, u) y_{I}(t, u) \differential{u} \eqcomma \\
         \left(\partial_t + \partial_s \right)y_{I}(t,s) &{} = - \gamma(s) y_{I}(t,s) \eqcomma \\
         \left(\partial_t + \partial_s \right)y_{R}(t,s) &{} = 0 \eqcomma 
    \end{aligned}
    \label{eq:limiting_pde_1}
    \end{align}
    with boundary conditions 
    \begin{align}
        y_S(t,0) & {} = 0 \eqcomma \nonumber \\
        y_I(t,0) & {} = \int_0^\infty y_{S}(t,s) \int_0^\infty \beta(s, u) y_{I}(t, u) \differential{u} \differential{s} \eqcomma \nonumber \\
        y_R(t,0) &{} = \int_0^\infty \gamma(s) y_{I}(t,s) \eqcomma 
    \end{align}
    and initial conditions $y_S(0, \bullet) : \setOfPositiveReals \to \setOfPositiveReals$, $y_I(0, \bullet) : \setOfPositiveReals \to \setOfPositiveReals$ such that 
    \begin{align}
        \int_0^\infty y_S(0,s) \differential{s} = 1 \eqcomma \quad \int_0^\infty y_{I}(0,s) \differential{s} = \rho \eqstop  
    \end{align}
    We set $y_R(0,s) = 0$ for all $s \in \setOfPositiveReals$ in keeping with our assumption that initially there are no recovered individuals. One can interpret $y_S(t,s), y_I(t,s)$, and $y_R(t,s)$ as the densities at time $t$ of susceptible, infected and recovered individuals at age $s$. 
    
    The limiting system of \acp{PDE} in \eqref{eq:limiting_pde_1} is linear in $y_S, y_I$, and $y_R$, but non-local. For different choices of the functions $\beta$ and $\gamma$ depending on the particular infectious disease in question, one can solve \eqref{eq:limiting_pde_1} numerically and fit to data. Typically, one would assume a parametric representation of the functions $\beta$ and $\gamma$ and then, attempt to infer those parameters based on data. However, a common problem in epidemiological literature is that the choice of the likelihood function is often ad hoc and strictly speaking, unjustifiable. To this end, the \ac{DSA} method \cite{KhudaBukhsh2019SDS,OSU_whitepaper,KhudaBukhsh2022Hospital}  provides, in a principled way, a  likelihood function based on a random sample of transfer times\footnote{We treat the infection time as a transfer time from the susceptible to the infected compartment. Similarly, the recovery time is seen as a transfer time from the infected to the recovered compartment.}. In the next section, we describe the \ac{DSA} method in greater detail. 
    
    
    \section{Parameter inference using \acl{DSA}}
    \label{sec:dsa}
    The \ac{DSA} method combines dynamical systems theory and survival analysis. For a given dynamical system, typically described by \acp{ODE} or \acp{PDE} for population counts/proportions,  the \ac{DSA} method provides an alternative interpretation that characterizes probability laws of transfer times \cite{KhudaBukhsh2019SDS,OSU_whitepaper}. The mathematical underpinning is provided by a novel application of the  {Sellke construction}. For the sake of simplicity, we assume in the following that the function $\beta(u,v)$ depends only on the age $v$ of the infected individual  and not on the age $u$ of the susceptible individual, \ie, $\beta(u,v) = \beta(v)$. This will allow for a simpler and a more intuitive description of the Sellke construction. 
    
    \subsection{Sellke construction}
    \label{sec:sellke}
    The classical Sellke construction \cite{anderson_britton} provides an alternative individual-based description of the standard stochastic mass-action \ac{SIR} epidemic model. It can be shown that the resultant epidemic process is equivalent to the original population-level stochastic model in the sense that the counts of individuals with different immunological statuses have the same probability law under both constructions. However, the crux of the Sellke construction is that it describes the epidemic process in terms of individual survival probabilities (\ie, for an initially susceptible individual, the probability of remaining susceptible till time $t$). This is useful for parameter inference. The classical Sellke construction can be adapted to the age-structured epidemic model of ours in a straightforward fashion. 
    
\begin{figure}[t]
        \centering
        \includegraphics[scale=0.75]{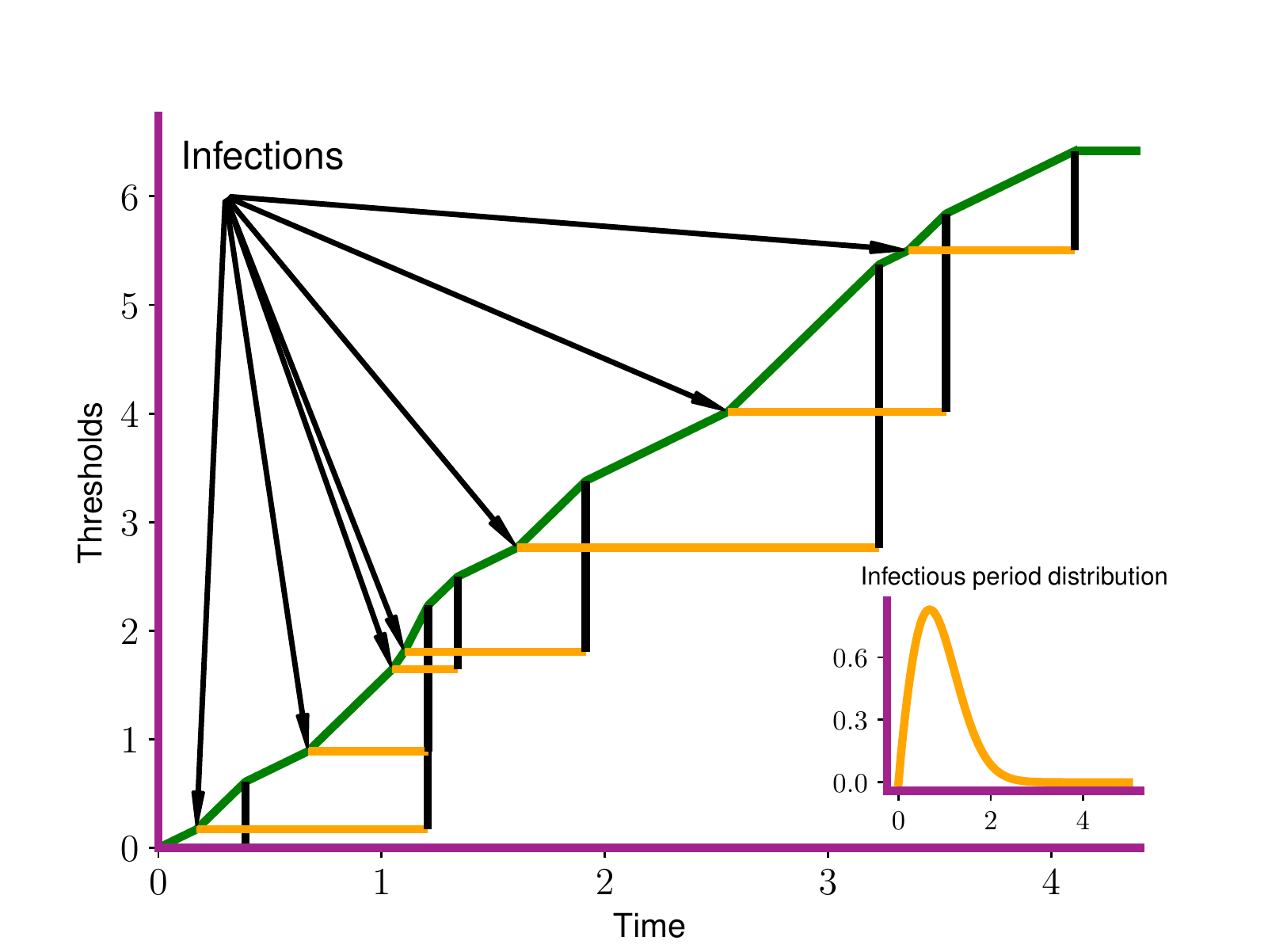}
        \caption{Sellke construction. Here, we begin with a single infected individual. 
        The arrows point to the times of infection.
        The orange horizontal lines indicate the infectious period of each infected individual. The \ac{PDF} of the infectious periods is shown in the inset (Weibull with shape $c=1.9$ and scale $1$). }
        \label{fig:sellke}
    \end{figure}

    As described in \Cref{sec:model}, suppose we begin with $n$ susceptible and $m$ infected individuals. To each of those $n$ susceptible individuals, we assign a threshold, an exponentially distributed random variable with mean one. Let $U_i$ denote the threshold corresponding to the $i$-th susceptible individual. The random variables $U_1, U_2, \ldots, U_n$ are independent. Let $U_{(1)}, U_{(2)}, \ldots, U_{(n)}$ be the corresponding order statistics, \ie, $U_{(1)}\le U_{(2)} \le \ldots \le U_{(n)}$. Let us now define the \emph{cumulative infection pressure}
    \begin{align}
        \mathsf{A}(t) \defeq \int_0^t \measureIntegral{ X_u^{I}  }{ n^{-1} \beta }  \differential{u} \eqcomma 
    \end{align}
    where the intensity function $\beta$ depends only on the age of the infected individuals. The epidemic process proceeds as follows: The first infection occurs when the cumulative infection pressure exceeds the smallest individual threshold, \ie, when $\mathsf{A}(t) \ge U_{(1)}$ for the first time; the second infection occurs when $\mathsf{A}(t) \ge U_{(2)}$, and so on. Note that infected individuals recover following an infectious period that has a probability law characterized by the hazard function $\gamma$. Therefore, it is possible that the cumulative infection pressure becomes constant when the last infected individual recovers and there are no more infected individuals. Susceptible individuals whose thresholds are never exceeded by the cumulative infection pressure $\mathsf{A}(t)$ escape infection and never leave the susceptible compartment. \cref{fig:sellke} provides a pictorial description of the Sellke construction. The resultant epidemic process, captured by measure-valued processes, is equivalent to the one described in \Cref{sec:model} (with the adjustment $\beta(u, v) =\beta(v)$).

    Let us denote the time of infection of an initially susceptible individual by $T_{I}$. In essence, the Sellke construction specifies an individual-level survival function: The probability that an initially susceptible individual $i$ remains susceptible till time $t$, conditional on the history (filtration) $\mathcal{H}_{t-}$ of the epidemic process, is given by 
    \begin{align}
        \probOf{T_{I} > t \mid \mathcal{H}_{t-} } = 
           \probOf{ U_i > \mathsf{A}_t \mid \mathcal{H}_{t-}  } = \myExp{- \mathsf{A}_t} =  \myExp{- \int_0^t \measureIntegral{ X_u^{I}  }{ n^{-1} \beta }  \differential{u}  } \eqcomma
           \label{eq:sellke_probability}
    \end{align}
    where $U_i \sim \distribution{Exponential}{1}$ is the threshold of the individual $i$. This survival probability will play a crucial role in devising the \ac{DSA}-likelihood function. It is worth pointing out that the random variable $T_I$ is improper because some individuals may escape infection with positive probability. 
    
    As $n \rightarrow \infty$, the scaled stochastic process $n^{-1}X_t$ converges to the vector-valued deterministic, continuous function $x_t \defeq (x_t^S, x_t^I, x_t^R)$. As before, the densities $y_S(t, \bullet), y_I(t, \bullet)$, and $y_R(t, \bullet)$ of $x_t^S, x_t^I$, and $x_t^R$ satisfy the system of \acp{PDE} given in \eqref{eq:limiting_pde_1}. Because of our assumption $\beta(u, v) = \beta(v)$, it makes sense to integrate out the age component  for the susceptible and the recovered individuals. Therefore, by defining 
    \begin{align*}
        z_S(t) \defeq \int_0^\infty y_S(t, s) \differential{s}, \text{ and } z_R(t) \defeq \int_0^\infty y_R(t, s) \differential{s} \eqcomma  
    \end{align*}
    we can write the limiting system as follows:
    \begin{align}
        \begin{aligned}
            \timeDerivative{z_S(t)} &{}  = - z_S(t)  \int_0^\infty \beta(s) y_I(t, s) \differential{s} \eqcomma \\
            \left(\partial_t + \partial_s \right) y_I(t,s) &{} = - \gamma(s) y_I(t,s) \eqcomma \\
            \timeDerivative{z_R(t)} &{}  = \int_0^\infty \gamma(s) y_I(t,s) \differential{s} \eqcomma
        \end{aligned}
        \label{eq:sellke_pde}
    \end{align}
    with initial conditions $ z_S(0) = 1, z_R(0) = 0 $ and $y_I(0, \bullet) : \setOfPositiveReals \to \setOfPositiveReals$ such that $$\int_0^\infty y_I(0, s) \differential{s} = \rho,$$  and boundary condition 
    \begin{align}
        y_I(t, 0) = z_S(t) \int_0^\infty \beta(s) y_I(t, s) \differential{s} \eqstop 
        \label{eq:sellke_pdeboundary}
    \end{align}
    Rewriting \eqref{eq:sellke_pde} and with the initial condition $z_S(0)=1$, we immediately see 
    \begin{align*}
        z_S(t) = \myExp{ - \int_0^t \int_0^\infty \beta(v) y_I(u, v) \differential{v} \differential{u} } = \myExp{-\int_0^t \measureIntegral{x_s^I}{ \beta} \differential{s}} \eqcomma 
    \end{align*}
    which is precisely the limit of the survival function $\probOf{T_I > t}$ in \eqref{eq:sellke_probability} as $n \rightarrow \infty$. Therefore, the function $z_S$, the limiting proportion of susceptible individuals, can be interpreted as a survival function. However, the survival function $z_S$ is improper because $z_S(\infty) > 0$. The quantity $z_S(\infty)$ is precisely the limiting proportion of susceptible individuals (who escape the infection). However, the survival function $z_S$ can be made proper by conditioning on individuals who get infected \cite{KhudaBukhsh2019SDS}. Another important observation is that the ``time to infection'' random variables associated with the initially susceptible individuals become independent in the limit of $n \rightarrow \infty$. This phenomenon is sometimes referred to as \emph{mean-field independence} \cite{Baladron2012Mean_field,Meleard1996Asymptotic}. 
    
    Furthermore, using the method of characteristics on \eqref{eq:sellke_pde}, we get 
    \begin{align*}
        y_I(t, s) &{} = \left\{  \begin{array}{cc}
            y_I(0, s-t) S_{\gamma}(s) / S_{\gamma}(s-t) \eqcomma & \text{ for } s>t \eqcomma \\
            y_I(t-s, 0) S_{\gamma}(s) \eqcomma  &  \text{ for } t \ge s \eqcomma
        \end{array} 
        \right.
    \end{align*}
    where $S_\gamma$ is the survival function of the probability distribution characterized by the hazard function $\gamma$. That is, $S_\gamma (t) = \myExp{- \int_0^t \gamma(s) \differential{s}}$. Unfortunately, $y_I$ does not admit an explicit solution. However, efficient numerical methods exist. We describe the solution scheme we adopted in \cref{appendix:pde_scheme}. The limiting proportion of recovered individuals $z_R$ is also fully described by the limiting density $y_I$ of infected individuals
    \begin{align*}
        z_R(t) = \int_0^t \int_0^\infty \gamma(v) y_I(u, v) \differential{v} \differential{u} = \int_0^t \measureIntegral{x_s^I}{\gamma} \differential{s} \eqstop 
    \end{align*}
    
    \subsection{Likelihood contribution of infection times}
    Let us denote by $\theta$ the set of parameters required to describe the contact interval distribution in terms of $\beta$ and the infectious period in terms of $\gamma$. On account of the Sellke construction, we can treat the function $z_S$ as an improper survival function for the (improper) random variable $T_I$, the time to infection for an initially susceptible individual. Therefore, we can define the conditional \ac{PDF}  
    \begin{align}
        f_{T, \theta}(t) \defeq  - \frac{1}{ \tau_{T} } { \timeDerivative{z_S(t)} } = \frac{ z_{S}(t) \measureIntegral{x_t^I}{\beta} }{ \tau_{T}} \eqcomma 
        \label{eq:infection_density}
    \end{align}
    for the infection times, where $\tau_T \defeq 1- z_S(T)$. Also, set $\tau \defeq \tau_\infty$. The \ac{PDF} $f_{T}$ is proper by virtue of the conditioning. 
    
    Most epidemic and pandemic trajectories are only partially observed. A crucial advantage of the \ac{DSA} approach is that it does not require the whole trajectory. Suppose we have a random sample of infection times $t_1, t_2, \ldots, t_K$ from an epidemic trajectory observed partially till time $T$, for some finite, positive number $T$. Then, following the mean-field independence, the contribution of the infection times to the \ac{DSA} likelihood function is given by 
    \begin{align}
        \ell_{I}\left(\theta  \right) & \defeq \prod_{i=1}^{K} f_{T, \theta}(t_i)  \eqstop 
        \label{eq:likelihood_I}
    \end{align}
    The contribution $\ell_{I}$ can be modified in a straightforward fashion if the infection times are censored and/or truncated.

    \subsection{Likelihood contribution of recovery times}
    Now, let us describe the contribution of the recovery times to the \ac{DSA} likelihood. While the recovery times are often not observed, or only partially observed (with further possibility of censoring or truncation), when available they can be incorporated into the \ac{DSA} likelihood function rendering it more informative. There are two possible scenarios. Let us consider the simpler case first: We have a random sample $s_1, s_2, \ldots, s_L$ of infectious periods. Then,  denoting the \ac{PDF} of the probability law characterized by the hazard function $\gamma$ by $r_{\gamma}$,  the contribution of the random sample of infectious periods to the \ac{DSA} likelihood function is given by 
    \begin{align}
        \ell_{R}^{(1)}\left( \theta  \right) & \defeq \prod_{i=1}^{L} r_{\gamma}(s_i) \eqstop 
        \label{eq:likelihood_R1}
    \end{align}
    Now, let us consider the second case: We do not directly  observe individual infectious periods, but only observe recovery times. Suppose $u_1, u_2, \ldots, u_M$ is a random sample of recovery times of $M$ individuals whose infection times are unknown. They are precisely a random sample of the sum of two independent random variables: Time to infection and infectious period. Therefore, we can define the convolution-form  \ac{PDF} 
    \begin{align}
        g_{T, \theta} (t) &{} \defeq \frac{g(t)}{ \int_0^T g(s) \differential{s} } \eqcomma 
        \label{eq:recovery_density}
    \end{align}
    conditional on the partially observed epidemic trajectory till time $T$, where
    \begin{align}
        g(t) &{} \defeq \int_0^t f_{T, \theta}(u) r_{\gamma}(t-u) \differential{u} \eqstop 
    \end{align}
    Now, with the conditional \ac{PDF} of the recovery times given in \eqref{eq:recovery_density}, we can write down the contribution of the random sample  $u_1, u_2, \ldots, u_M$ of recovery times as follows
    \begin{align}
        \ell_{R}^{(2)} \left( \theta  \right) & {} \defeq \prod_{i=1}^{M} g_{T, \theta}(u_i) \eqstop 
                \label{eq:likelihood_R2}
    \end{align}
    The conditional \ac{PDF} $g_{T, \theta}$, in general, does not admit a closed-form expression. However, it can be computed numerically. 
    
    \subsection{The \ac{DSA} likelihood}
    Suppose we have a random sample  $t_1, t_2, \ldots, t_K$ of infection times, a random sample $s_1, s_2, \ldots, s_L$ of infectious periods, and a random sample $u_1, u_2, \ldots, u_M$ of recovery times. Then, the \ac{DSA} likelihood function is given by 
    \begin{align}
        \ell\left( \theta \right) \defeq \ell_{I} \left( \theta  \right)  \times  \ell_{R}^{(1)} \left( \theta  \right) \times  \ell_{R}^{(2)} \left( \theta  \right) \eqstop 
        \label{eq:DSA_likelihood}
    \end{align}
    Note that it is not necessary to have data on recovery times. The likelihood contribution $\ell_{I} \left( \theta  \right)$ is adequate for parameter inference. See \cite{OSU_whitepaper} where parameter inference was done for the COVID-19 pandemic in the state of Ohio, USA based only on infection times. When information on recovery times are unavailable, we simply set $\ell_{R}^{(1)}=1$ and $\ell_{R}^{(2)} = 1$ by adopting the convention $\prod_{i=1}^{0} s_i = 1$. 
    
    Often it is easier to work with the log-likelihood function. Therefore, for the purpose of parameter inference, we also define the \ac{DSA} log-likelihood function 
    \begin{align}
        \mathcal{L} \left(\theta \right) \defeq \log( \ell\left( \theta \right)) = \log(\ell_{I} \left( \theta  \right) ) + \log(\ell_{R}^{(1)} \left( \theta  \right) ) + \log(\ell_{R}^{(2)} \left( \theta  \right) )\eqstop  
        \label{eq:DSA_loglikelihood}
    \end{align}
    The \ac{MLE} $\hat{\theta}$ of the parameter $\theta$ is then numerically obtained by maximizing the log-likelihood function $\mathcal{L}\left(\theta\right)$. That is, 
    \begin{align}
        \hat{\theta} \defeq \arg\max_{\theta} \mathcal{L}\left(\theta\right) \eqstop 
    \end{align}
    We present numerical results in \cref{sec:numerical}.  For Bayesian methods, we need to introduce a prior for the parameter $\theta$ and then implement a \ac{MCMC} algorithm to approximate the posterior distribution of the parameter $\theta$. We, however, do not pursue the Bayesian path in this paper.  
    
    \subsection{Mean-field limits as Chapman--Kolmogorov equations}
    An alternative way to view \ac{DSA} is to interpret the limiting trajectory equations as satisfying Chapman--Kolmogorov equations (written in the differential form) for certain probability distributions. Let us pick a random individual in the (infinitely large) population and follow in time. Let $W(t) \in \{ \mathsf{S}, \mathsf{I}, \mathsf{R} \}$ denote a Markov process that keeps track of the immunological status of the individual. Write $p_t \defeq (p_t^{S}, p_t^I, p_t^R)$ for $p_t^\star \defeq \probOf{W(t) = \star}$. Then, following the previous discussion, \ac{DSA}, in essence,  is tantamount to writing  
    \begin{align}
        \begin{aligned}
            p_t^{S} = {} \frac{z_S(t)}{1+\rho}\eqcomma\quad  p_t^{I} = {} \frac{z_I(t)}{1+\rho}\eqcomma\quad   p_t^{R} = {} \frac{z_R(t)}{1+\rho}\eqcomma 
        \end{aligned}
    \end{align}
    where $z_I(t) = \int_0^\infty y_I(t, s) \differential{s}$. It is in this viewpoint that we say the limiting mean-field equations given in \cref{eq:limiting_pde_1} satisfy the Chapman--Kolmogorov equations for the probability distribution $p_t$. It is worth mentioning that the time derivative $\timeDerivative{p_t}$ gives us what is popularly known as the \ac{CME} in the physical sciences literature. 
    
    \subsection{Estimate of effective population size}
    In addition to giving a simple product-form likelihood function for $\theta$, \ac{DSA} also gives a ready estimate of the effective population size. Given $k_T$, the number of cases observed by time $T$, the effective population size can be estimated by the discount estimator 
    \begin{align}
        \hat{n}_T \defeq \frac{k_T}{1- z_S(T)} \eqstop 
        \label{eq:effective_popln_size}
    \end{align}
    In similar vein, we can also estimate the final size of the epidemic as follows
    \begin{align}
        \hat{k}_{\infty} = \frac{\tau k_T }{1- z_S(T)} \eqstop  
        \label{eq:final_size}
    \end{align}
    Please refer to \cite{KhudaBukhsh2019SDS, OSU_whitepaper} for further discussions on this. 
    
    \section{Numerical results}
    \label{sec:numerical}
    In this section, we demonstrate how the \ac{DSA} method can be used for inference of model parameters from  infectious disease outbreak data using the likelihood functions described in~\Cref{sec:dsa}. Typical outbreak data consist of population-level aggregated counts (such as the daily number of newly positive cases). Hence, we use this scenario as a benchmark for numerical validation. At the beginning, we will analyse synthetic data and make several simplifying assumptions, which we will gradually remove in favour of more realistic models when considering datasets from real epidemic outbreaks, such as the \ac{FMD} and the COVID-19 pandemic in India. 

    \subsection{Synthetic data}
    We begin by carrying out \ac{DSA} analysis on synthetic data. We begin by keeping the premise deliberately simple: We assume the family of the infectious period is known in that the functional form of the hazard function $\gamma$ (or the \ac{PDF} characterized by $\gamma$) is known, but the parameters are to be inferred along with the initial condition of the \ac{PDE}~\eqref{eq:limiting_pde_1} and a constant infection rate, $\beta$. To this end, we begin by assuming the infectious period is a Gamma random variable. The rationale behind this choice is the flexibility of the Gamma distribution and its historical importance in infectious disease epidemiology \cite{vanmieghemNonMarkovianInfectionSpread2013,eichnerTransmissionPotentialSmallpox2003,krylovaEffectsInfectiousPeriod2013,wearingAppropriateModelsManagement2005, byrneInferredDurationInfectious2020}. The proposed inference scheme, of course, works for any other distribution, such as the log-logistic or Weibull (not reported here). All the code to reproduce the results in this section is available online~\footnote{\url{https://github.com/Zkeggia/DSA_refactor}}, and a brief description of the numerical scheme used to solve the \ac{PDE} can be found in \Cref{appendix:pde_scheme}.
    
    \paragraph{Description of data}
    The Sellke construction is an excellent means to generate exact simulations of an epidemic. We simulate an outbreak on a  population of $N=10000$ individuals. Epidemics are run until no infected individuals are present in the population. Datasets consist of the series of infection and recovery times taken from the simulation, without noise nor delays. 
    
    \begin{figure}[h!]
        \centering
       \includegraphics[width=0.4\textwidth]{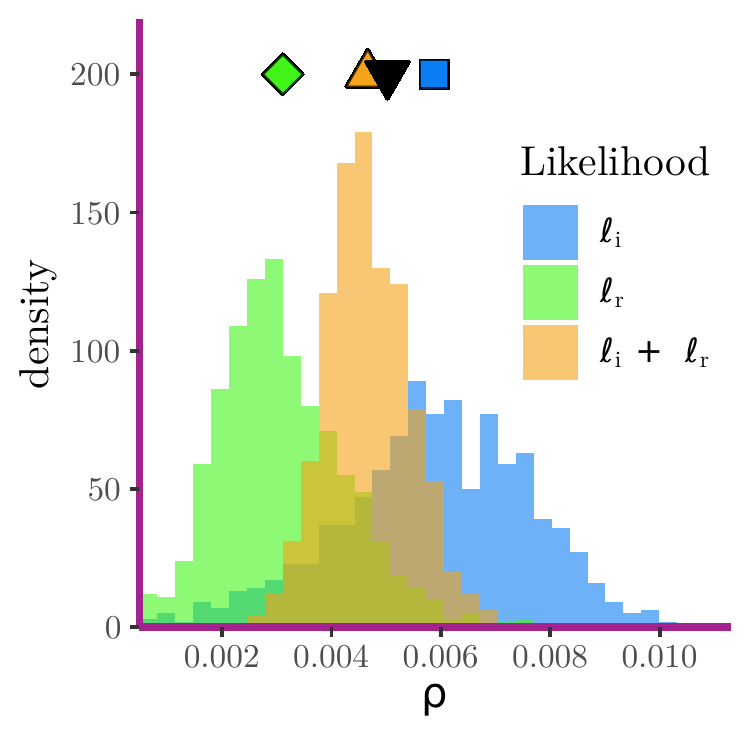}   
        \includegraphics[width=0.4\textwidth]{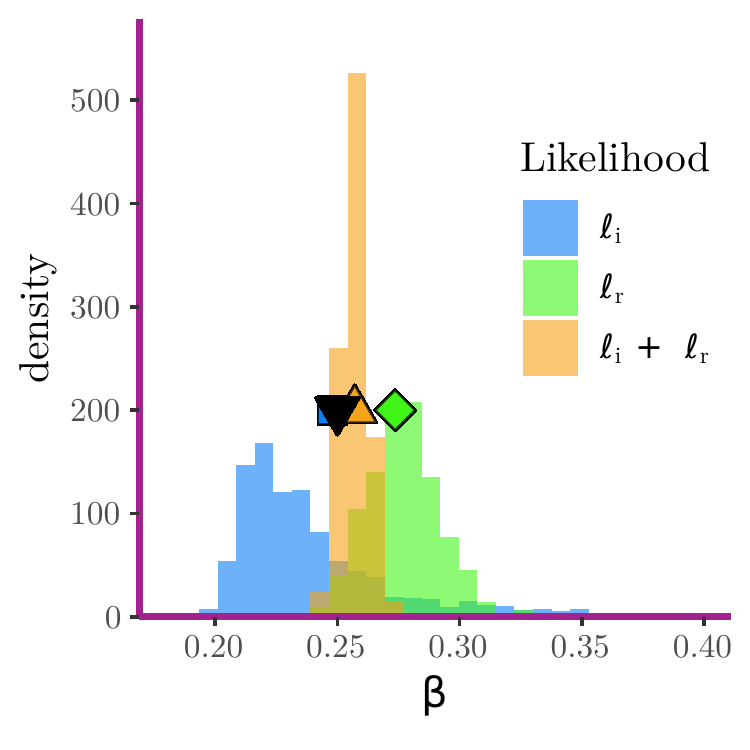}
        \caption{Inferred parameters $\rho$ (left), and $\beta$ (right). Each figure shows histograms for different scenarios of data availability, as denoted in the legend. The true parameter is represented by the downward  triangle. The square is the average value inferred when considering only infectious times, the diamond when considering only recovery times, and the upward triangle when considering both.}
        \label{fig:synthetic_inferred_1}
    \end{figure}

    We consider three different scenarios, characterised by different availability of data: we either work  with only recovery times,  with only infection times, or with both. We generate $1000$ datasets from the same initial conditions, to characterize the distribution of the estimates. Estimates are found by means of a mix of global and local optimization routines.
    
\begin{figure}
        \centering
        \includegraphics[width=0.8\textwidth]{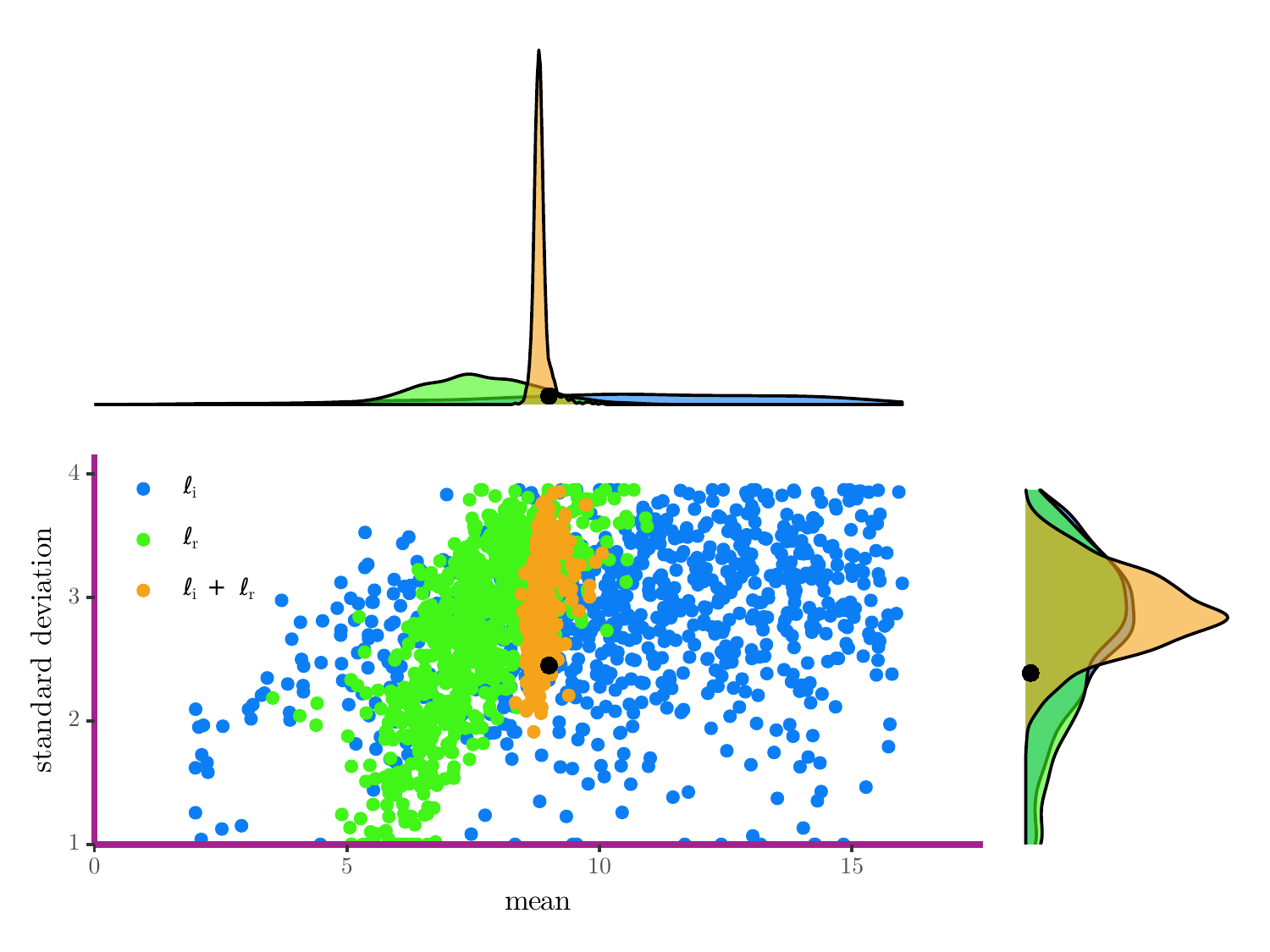}
        \caption{Inferred infectious period  distribution mean and standard deviation. Black dots represent the true values.}
        \label{fig:synthetic_inferred_2}
    \end{figure}

    The objective is to infer the initial proportion of infected individuals $\rho = 50/9950$, the per-contact infection rate $\beta=0.25$, and the parameters of the distribution of infectious period, which is a Gamma distribution with mean $\mu=9$ and variance $\sigma^2=6$. Results are shown in figures~\ref{fig:synthetic_inferred_1} and~\ref{fig:synthetic_inferred_2}. 
    
    We find that inference based on only infection times using the likelihood function $\ell_{I}(\theta)$ in \eqref{eq:likelihood_I} results in wider distributions for all inferred parameters, suggesting greater uncertainty, than inference based on both. This is expected because the likelihood function $\ell(\theta)$ in \eqref{eq:DSA_likelihood} is more informative than the likelihood function $\ell_{I}(\theta)$ in \eqref{eq:likelihood_I}. In general, the true parameters are always near the mode of the distributions of the inferred parameters. It is worth noting that when the infection rate $\beta$ is overestimated, the initial proportion of infected individuals $\rho$ is underestimated, and vice versa. This suggests a potential statistical unidentifiability of the parameters. Outbreaks starting with a higher number of infected individuals but smaller transmission rate may be hard to distinguish from those that start with a smaller number of infected individuals but with higher transmission rate. 


    The mean and the standard deviation of the distribution of the infectious period are reported in \Cref{fig:synthetic_inferred_2}. We observe that inference based only on infection times, in general, accurately captures the mean of the distribution of the infectious period but tends to overestimate the variance. The overall quality of inference improves significantly when recovery times are also available. 
    

    \subsection{\acl{FMD}}
    Let us now turn to real datasets. We consider the 2001 \ac{FMD} outbreak in the UK. The outbreak began in February 2001 and ended in September 2001, affecting more than $2000$ farms. The government efforts to control the epidemic resulted in the culling of millions herds and flocks~\cite{daviesFootMouthDisease2002}. Because of the specific interventions taken to control to outbreak, we interpret the infectious period in the \ac{DSA} model as the time from when the disease hit a farm to elimination of infected herds, \ie,  the time to removal. Since this quantity is unlikely to be exponentially distributed, we fir a  gamma distribution. For the contact interval distribution characterized by the hazard function $\beta$, we assume a Weibull distribution. 
    \begin{figure}
      \centering
      \includegraphics[scale=0.85]{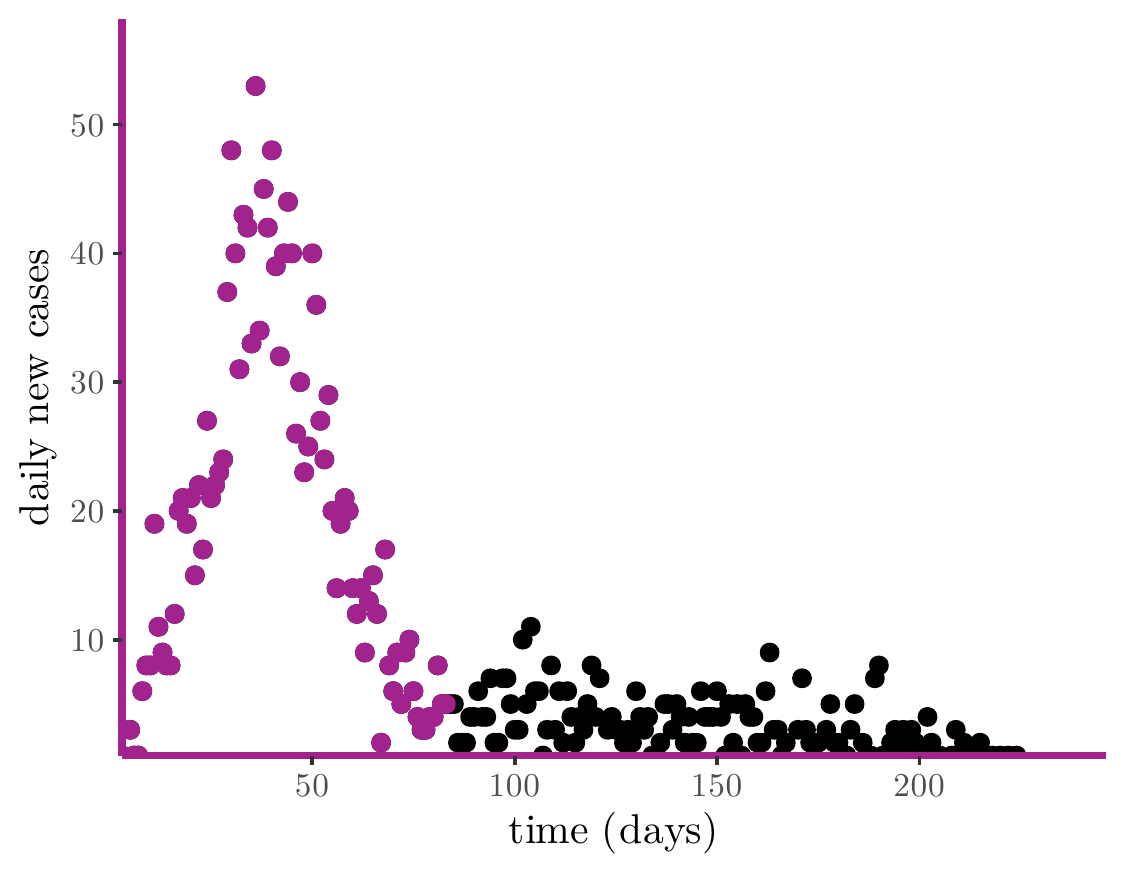}
      \caption{Visualisation of the \ac{FMD} outbreak. New daily cases since the first day of data (February 2001), to last day where a new case was confirmed (September 2001). The data points in black are excluded from the analysis.}
      \label{fig:foot_and_mouth_dataset}
    \end{figure} 
 
   The dataset\footnote{Data on daily incidence kindly provided by Professor Michael Tildesley, University of Warwick.} consists of daily incidence of infected premises by time of report, $\{t_i,I_i\}$, with no information on removal times. See \Cref{fig:foot_and_mouth_dataset}. For each day $t_i$, we distribute the number of new cases $I_i$  uniformly in the interval $(t_{i-1}, t_i)$. Furthermore, we consider only the first $80$ days of data, to exclude the noisy tail and potentially confounding effects of strict measures. This simplifying assumptions allows us to maximize the likelihood $\ell_I(\theta)$ in \eqref{eq:likelihood_I}.  Since the original data points are too noisy, we consider the 7-day moving average of the counts, starting from day $6$. This results in a smoother dataset that is less noisy, although a bit delayed with respect to the true one.
   
    \begin{figure}
        \centering
        \includegraphics[scale=0.8]{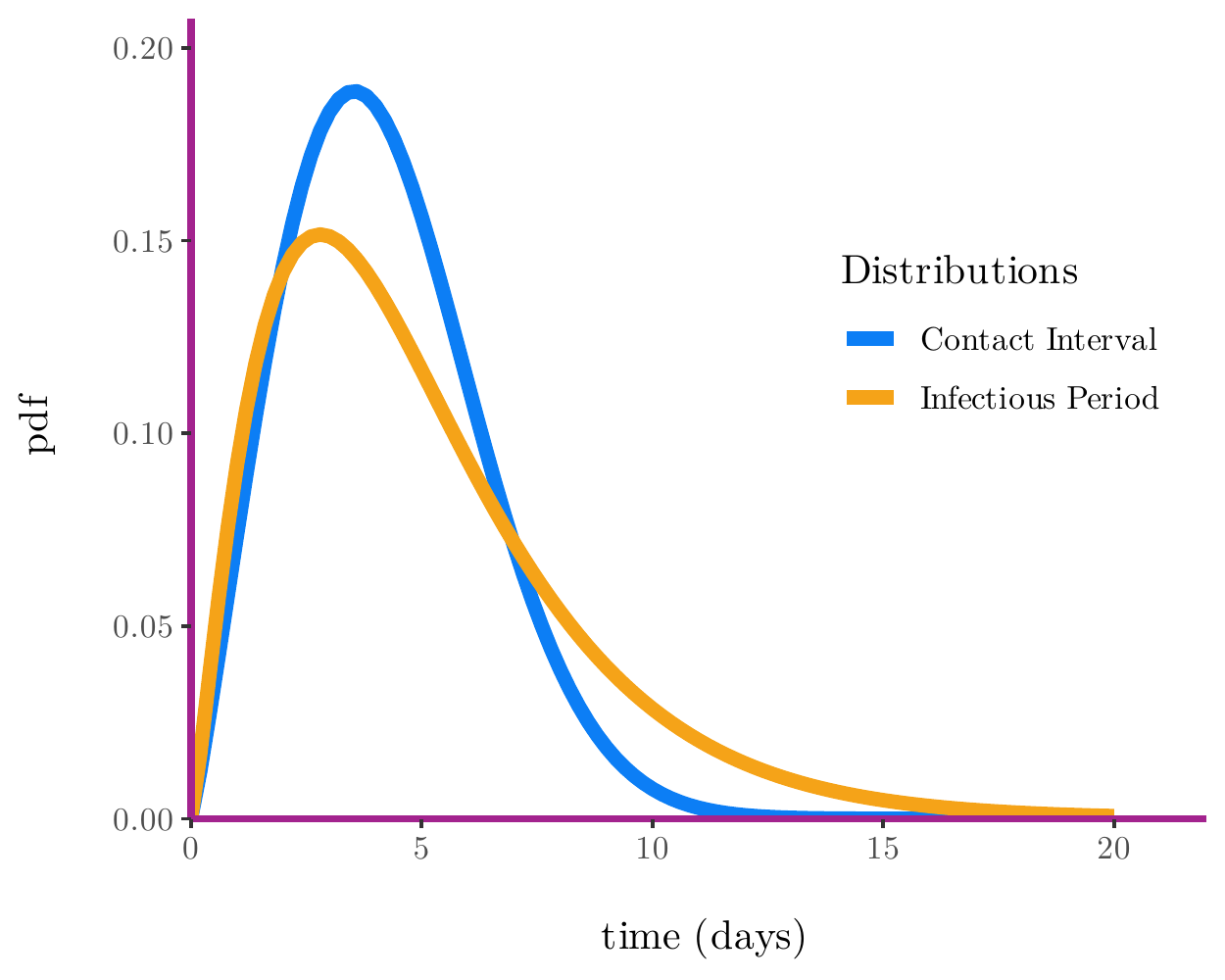}
        \caption{The best fitting \acp{PDF} of the contact interval and the infectious period inferred from the \ac{FMD}  data.}
        \label{fig:pdfs_foot_and_mouth}
    \end{figure}
    
    Maximum likelihood estimates are obtained by means of a mix of global and local optimization routines. The distributions of inferred contact interval and infectious period are shown in \Cref{fig:pdfs_foot_and_mouth}. These shapes of the inferred distributions are in line with findings from other studies of same outbreak~\cite{fergusonFootandMouthEpidemicGreat2001}. Our model with Weibull contact interval distribution and Gamma infectious period does not consider  the incubation period explicitly. Once both infectious period and contact interval distributions are known, we can find $R_0$ using the formula 
    $R_0 = \int_0^\infty S_{\gamma}(t) \beta(t) \differential{t}$~\cite{MA2020129}, where $S_{\gamma}$, we recall, is the survival function of the infectious period distribution. This gives a point-estimate of $R_0=2.55$. 

   We compute confidence intervals using a bootstrap method, which we describe now. We first solve the limiting \ac{PDE}~\eqref{eq:limiting_pde_1} with the \ac{MLE} estimates. From the solution, we compute the distribution of infection times \ac{PDF}~\eqref{eq:infection_density}. This distribution is used  to generate $n=500$ synthetic datasets with as many datapoints as the original one, consisting of simulated dates of infections, on which we repeat the inference. Each new set of inferred parameters is then used to produce both the estimate for $R_0$ (shown in \Cref{fig:R_0fmd}), and the $(t,I(t))$ incidence curve that we can compare against the true data.  
   

   \begin{figure}[t!]
    \centering
    \includegraphics[scale=1]{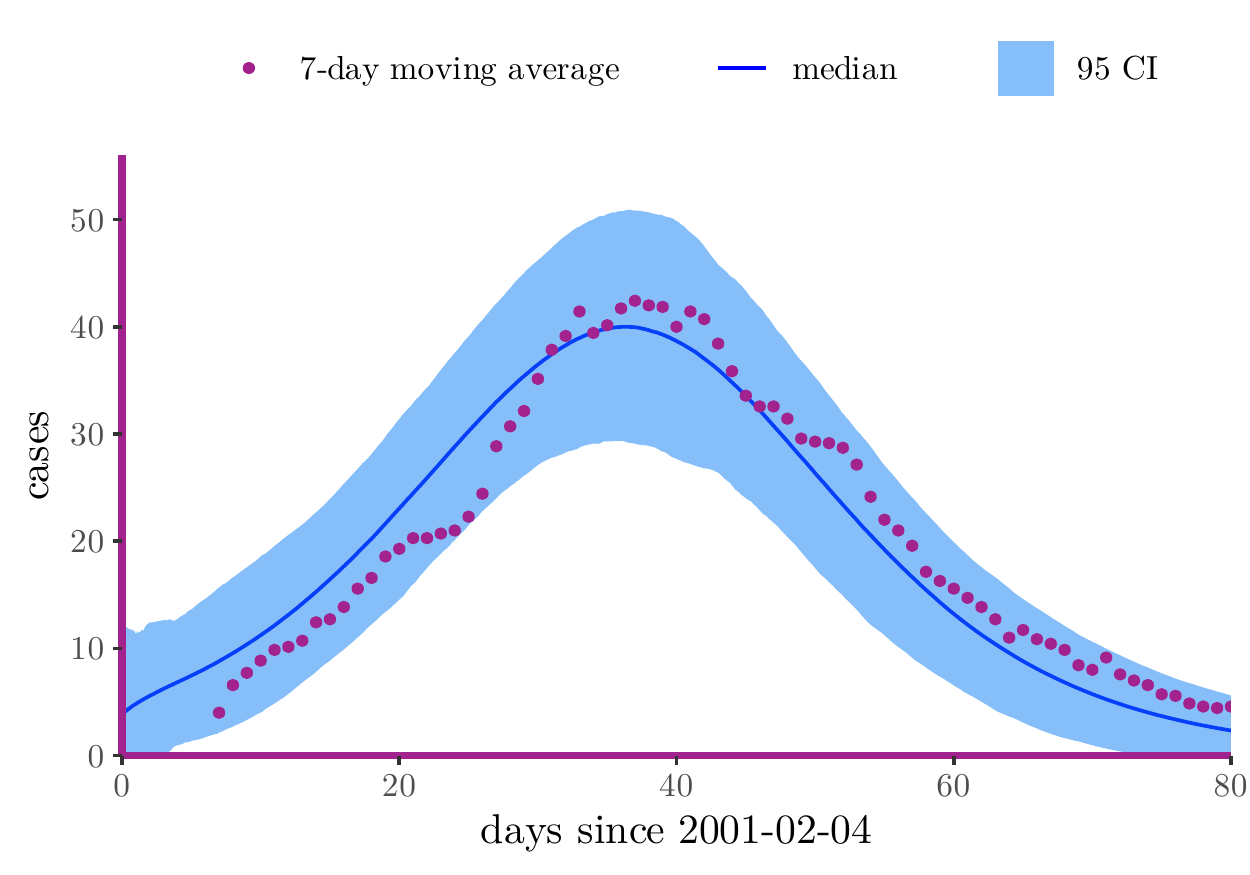}
    \caption{Variance-adjusted confidence intervals for the \ac{FMD} dataset.}
    \label{fig:foot_and_mouth_result}
    \end{figure}
    

   Finally, when computing confidence intervals, we compensate for other sources of noise that cannot be explicitly accounted for in our the model but are present in real-world data, such as testing limits, day-of-the-week effects, and various sources of delays. This variance-adjustment is done by inflating the confidence intervals by a factor determined by taking the square root of the variance between the data points and the 7-day moving average. Results are shown in \Cref{fig:foot_and_mouth_result}. As can be verified, the trajectories do capture the epidemic trend quite well in that all the data points lie within the variance-adjusted 95 Confidence Interval. 
   
    \subsection{Third wave of COVID-19 in India}
   \begin{figure}
      \centering
       \includegraphics[scale=1]{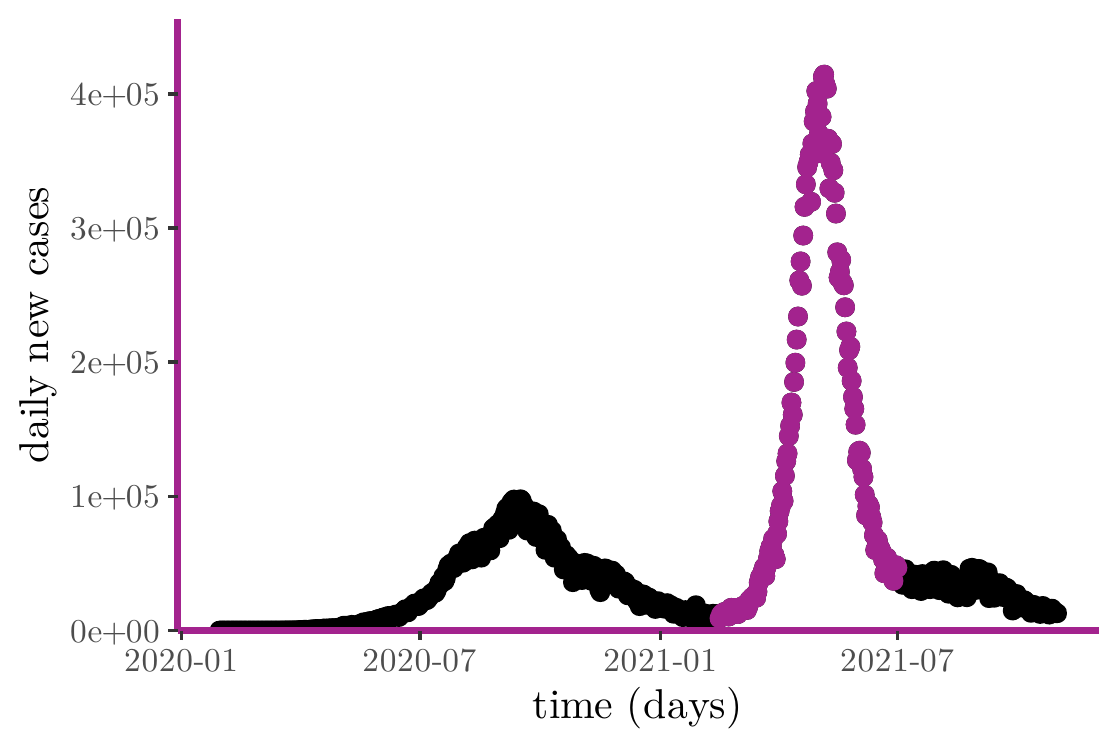}
       \caption{Indian wave of COVID-19 cases. The Delta-wave we fit the model to is highlighted in purple, and spans from 2021-02-15 to 2021-06-31.}
       \label{fig:India_cases}
   \end{figure}
   The analysis of \ac{FMD} outbreak data makes use of only infection times. As the synthetic data analysis suggests inference based only on infection times tend to be poorer compared to when both infection times as well as recovery times are available, we now analyse an epidemic where both times are available.  
   

   In a global effort to document and control the ongoing Covid-19 pandemic, many governments provided freely available population-level datasets that we can use as case studies for inference when both infection and recovery times are known. Various countries adopted strong non-pharmaceutical measures that drastically changed the local dynamics of the epidemic, resulting in several distinct epidemic waves. At the same time, new Sars-Cov-2 variants emerged with markedly different epidemiological characteristics. To curtail the impact of such exogenous factors, we consider only the third wave in India~\footnote{Data available at https://api.covid19india.org/documentation/csv/}. Data consist of daily incidence and prevalence of cases, recoveries and deaths, meaning that we have data to inform both likelihoods in \eqref{eq:likelihood_I} and~\eqref{eq:likelihood_R2}. The observed period spans from 15 February 2021 to 31 June 2021 included. See \Cref{fig:India_cases}. For this dataset, we assume both the contact interval and the infectious period to be gamma distributed. 


   Similar to our approach on the \ac{FMD} data, daily cases are distributed uniformly across the day. Because the \ac{DSA} method requires only a random sample infection and recovery times, we work with a dataset generated by taking a random sample (without replacement) of size $3000$. We do not consider exogenous factors such as under-reporting of cases as they are beyond the scope of this paper. It is worth noting, however, that these exogenous factors surely have an impact on the results and can be accounted by a more refined model.
   
\begin{figure}[h!]
    \centering
    \includegraphics[scale=0.8]{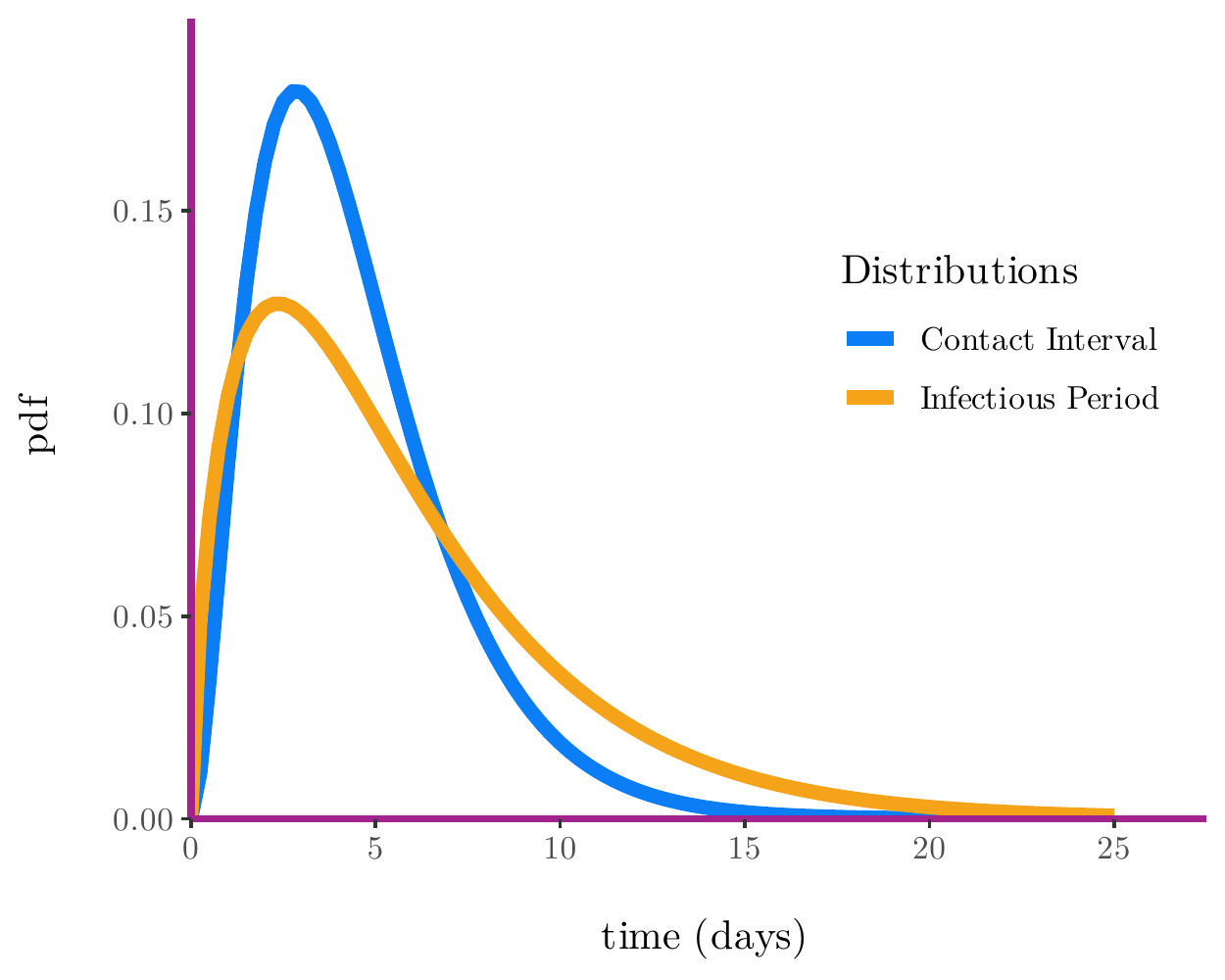}
    \caption{The best-fitting contact interval and infectious period distributions inferred from Indian Delta wave data. The distributions are, respectively, $\Gamma (4.5,10)$, and $\Gamma (5.5,20)$.}
    \label{fig:pdfs_indian}
\end{figure}

    The best-fitting inferred contact interval and infectious period  distributions are shown in \Cref{fig:pdfs_indian}. Ther are roughly in line with estimates of viral load and recovery distributions, respectively, from the literature~\cite{arroyo2021}. The point estimate for the reproduction rate is $R_0 = 1.69$. Although $R_0$ of Sars-Cov-2 Delta variant is estimated to be in the range $3-8$~\cite{liuReproductiveNumberDelta2021},  it is more realistic to compare our estimate with $R_t$ calculated from observed cases in that period, as our model uses only that source of information. The recovery distribution has a mean of $5.6$ days and a variance of $26$ days, so it is rather wide and right-skewed. The contact interval distribution is more peaked, with a slightly lower mean (around $4.5$ days) and a variance of roughly $10$. It is important to notice that infection times represent the collection of specimen from infected individuals, and recovery times follow country-specific healthcare system  protocols, so they do not necessarily coincide with the true infectious distributions. Furthermore, the infectious period start immediately after the incubation time has passed, while time to recovery is usually calculated from symptoms onset.   

\begin{figure}
        \centering
        \includegraphics{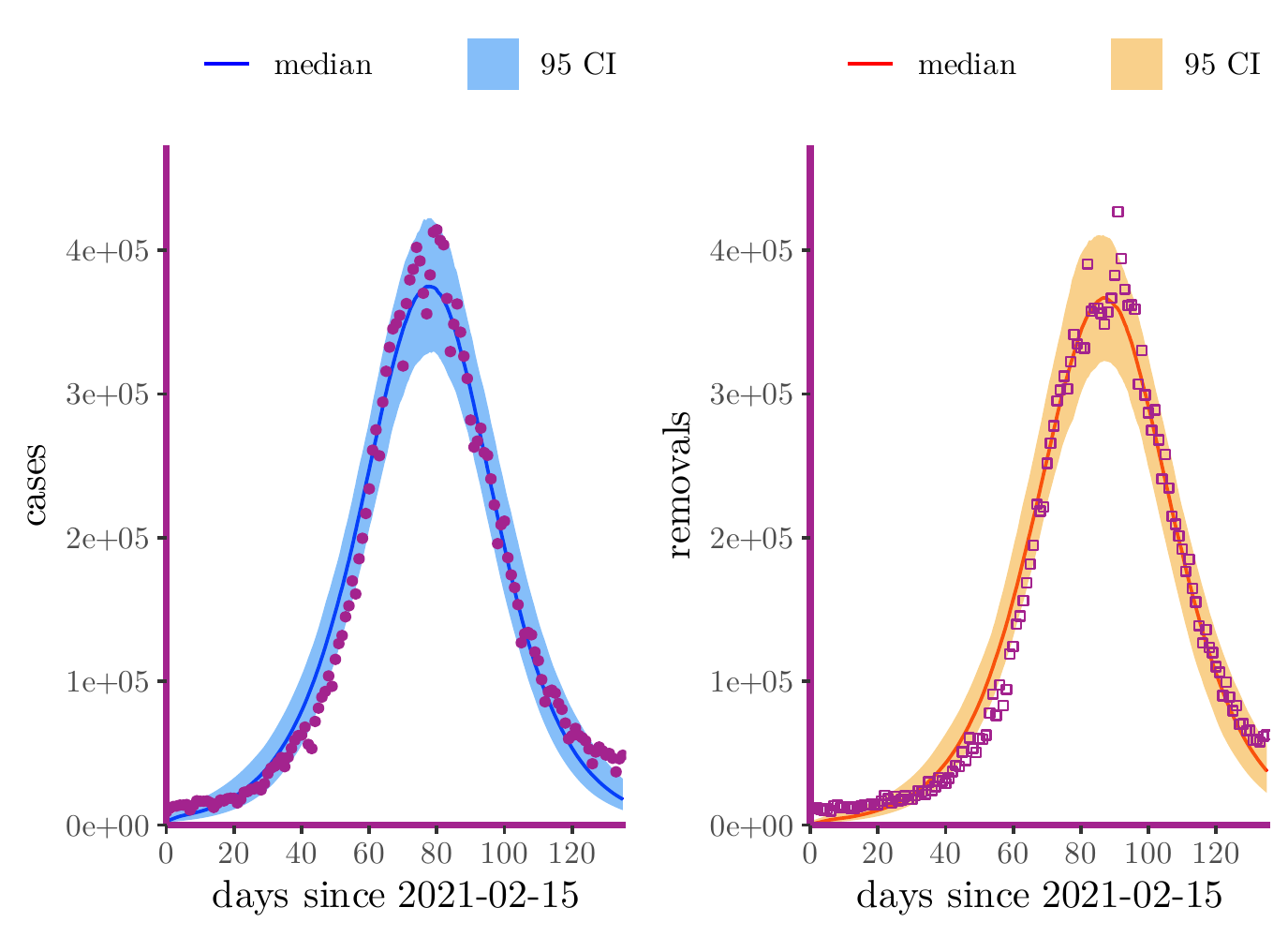}
        \caption{Confidence intervals for Indian wave. (Left) daily number of new cases, (right) daily number of recoveries or deaths (referred to as removals).}
        \label{fig:Indian_result}
    \end{figure}

    Confidence intervals are computed in a similar way to the \ac{FMD} analysis, with two major differences: The 7-day moving averages result in curve that is too delayed with respect to the actual one because of exponential growth/decline. Although this effect may be accounted for by considering exponential moving averages, we preferred not to modify the data that way. For a similar reason, computing the variance-adjusted confidence intervals that take into account all the noise that cannot be explained by the model is out of reach. Therefore, the confidence intervals, displayed in \Cref{fig:Indian_result}, underestimate the true variability of the underlying process, but seem to be generally in good agreement with the data. Interestingly, repeating the inference on different subsets of the original dataset, does not produce significantly different estimates for the two distributions of interest. This suggest that the method is robust, not only because we have many data points to inform the likelihood, but also because we consider both the infection times and the recovery/death times. The distribution of the estimates of the reproduction number is shown in the appendix (\Cref{fig:R_0India}).
      
    \section{Discussion}
    \label{sec:discussion}
    In this paper, we presented a  method called \ac{DSA} to both model and infer  parameters of non-Markovian epidemic models. A crucial advantage of \ac{DSA} is that it makes available the entire toolkit of survival analysis to making inference on  dynamical systems. Therefore, \ac{DSA} handles censored, truncated data in a straightforward and principled way. For instance, see \cite{KhudaBukhsh2021Prison} for an application of the \ac{DSA} method adapted to a simple Markovian \ac{SEIR} model where a snapshot of COVID-19 positivity data gathered through mass testing are used to analyse transmission in an Ohio prison. The analysis helped uncover the grave COVID-19  situation in correctional facilities in Ohio. Also, see \cite{KhudaBukhsh2022Israel} where we used the \ac{DSA} approach coupled with \ac{ABC} method to quantify the population-level effect of the mass vaccination campaign against COVID-19 in Israel. The analysis further helped quantify the indirect effect of vaccination on the unvaccinated young population in Israel. In \cite{Harley2022Ebola}, the \ac{DSA} method was used to analyse the individual-level epidemic data from the Ebola pandemic in the Democratic Republic of Congo suggesting success of the ring vaccination and contact tracing efforts evident from much lower estimate of the effective population size than previous analyses. 
    
    In this paper, we adopted the law of mass-action to model the interactions among the individuals for the sake of simplicity. Under the law of mass-action, an infected individual can potentially infect any susceptible individual in the population. This is in contract to network-based models where infected individuals can only infect their neighbors (connections defined by the graph adjacency matrix). However, inferring the underlying network structure is a nontrivial task and often infeasible. Therefore,  the mass-action models are still routinely used despite being unrealistic in many epidemics. Nevertheless, an immediate future direction for us would be to develop the \ac{DSA} methodology for a non-Markovian network model. 
    
    The crux of the \ac{DSA} methodology lies in the change in perspective about dynamical systems -- one that views them as describing probability distributions of times of infection and recovery, as opposed to describing (scaled) counts. As such, the method is completely general and could be quickly adapted to the particular setting of any infectious disease. We hope the software package \cite{GitHubLink} will help translate the \ac{DSA} methodology into a useful practical tool in modern infectious disease epidemiology. 

    \appendix
    \renewcommand{\theequation}{\Alph{section}.\arabic{equation}}
    
    \section{Brief derivation of the mean-field limit}
    \label{appendix:derivation}
    We provide an intuitive derivation of the \ac{PDE} limit discussed in \cref{sec:model} for the scaled stochastic process $n^{-1}X_t$. The proof follows a standard line of argument via the tightness-uniqueness route for Banach space-valued Markov processes. Similar (and more elaborate) derivations can be found in \cite{Fournier2004microscopic,Champagnat2008individual,Tran2008limit,Tran2009traits,Meleard:2012:SFS}. For the sake of completeness, we furnish a short overview of the main arguments here. 
    
    \subsection{Trajectory equations}
    In order to write down the trajectory equations for the components of $X_t$, we need to fix a partial order on the ages so as to make statements such as ``age of the $i$-th individual'' unambiguous. Let us fix the ``greater than or equal to'' relation on $\setOfPositiveReals$. Now, for  $i = 1, 2, 3, \ldots$,  we define maps $\sigma_i:  \mathcal{M}_{P}(\setOfPositiveReals) \rightarrow \setOfPositiveReals$,  which gives us the age of the $i$-individual (\ie, the $i$-th atom of a finite, point measure). Therefore, $\sigma_{i}(X_t^{I})$ is the age of the $i$-th infected individual at time $t$. In order to describe the interactions, we shall assume the stochastic law of mass action. Now, assuming there are only susceptible and infected individuals initially, we can write down the trajectory equations for the measure-valued stochastic processes $X_t^S, X_t^I$, and $X_t^R$ as follows:
    \begin{align}
        \begin{aligned} 
        X_t^{S} &{} = \sum_{k=1}^{ N_{S}(0) } \delta_{ t + \sigma_{k}(X_{0}^{S})  } \\
        &{} \quad - \int_0^t  \int_{\setOfNaturals} \int_0^\infty \delta_{t-s + \sigma_{i}(X_{s-}^{S}) } \indicator{i \le N_{S}(s-)} \indicator{\theta \le  \measureIntegral{X_{s-}^{I}}{ n^{-1} \beta(\sigma_{i}(X_{s-}^{S}), \bullet) }  } Q_1(\differential{s}, \differential{i}, \differential{\theta} ) \eqcomma \\
         X_t^{I} &{} = \sum_{k=1}^{ N_{I}(0) } \delta_{ t + \sigma_{k}(X_{0}^{I})  } + \int_0^t  \int_{\setOfNaturals} \int_0^\infty \delta_{t-s  } \indicator{i \le N_{S}(s-)} \indicator{\theta \le \measureIntegral{X_{s-}^{I}}{ n^{-1} \beta(\sigma_{i}(X_{s-}^{S}), \bullet) }  } Q_1(\differential{s}, \differential{i}, \differential{\theta} ) \\
         &{} \quad - \int_0^t \int_{\setOfNaturals} \int_0^\infty \delta_{t-s + \sigma_{k}(X_{s-}^{I}) } \indicator{i \le N_{I}(s-)} \indicator{\theta \le \gamma( \sigma_{i}(X_{s-}^{I})) } Q_2(\differential{s}, \differential{i}, \differential{\theta}) \eqcomma \\
         X_t^{R} &{}=  \int_0^t \int_{\setOfNaturals} \int_0^\infty \delta_{t-s } \indicator{i \le N_{I}(s-)} \indicator{\theta \le \gamma ( \sigma_{i}(X_{s-}^{I})) } Q_2(\differential{s}, \differential{i}, \differential{\theta}) \eqcomma
        \end{aligned}
        \label{eq:trajectory_equation}
    \end{align}
    where $Q_1$, and $ Q_2$ are independent \acp{PPM} with intensity measures $\differential{s}\times \differential{i}\times \differential{\theta}$ with Lebesgue measures $\differential{s}, \differential{\theta}$ on $\setOfPositiveReals$ and a counting measure $\differential{i}$ on $\setOfNaturals$. The \ac{PPM} $Q_1$ keeps track of infectious contacts, while the \ac{PPM} $Q_2$ book-keeps the natural recoveries of infected individuals.  The intensity function $\beta$ is scaled by a factor of $n^{-1}$ following the stochastic law of mass action \cite{anderson_britton,anderson_kurtz_CRN}. 
    
    \subsection{Assumptions}
     It is sufficient to assume that the global jump rates  (in terms of the instantaneous intensity functions $\beta$ and $\gamma$) of the Markov process $X_t$ are bounded above by a positive, finite quantity and that the initial population size does not explode in the sense that $\Eof{n^{-1}(N_S(0)+N_{I}(0))} < \infty$ in order to ensure the trajectory equation \eqref{eq:trajectory_equation} admits a unique path-wise solution $(X_t^{S}, X_t^{I}, X_t^{R})$. This follows from arguments similar to \cite[Theorem~2.5]{Tran2008limit} (see also \cite{Fournier2004microscopic,Tran2009traits,KhudaBukhsh2020Delay}). To see this, note that trajectories satisfying \eqref{eq:trajectory_equation} can be simulated by means of a straightforward adaptation of the Doob--Gillespie algorithm, which can be summarized as follows: i) Given an initial condition satisfying the technical assumptions, compute the next event time (either an infection or a recovery) by drawing an exponential random variable with rate equal to the global jump rate (total hazard) $\left( \int \measureIntegral{X_t^{I}}{n^{-1} \beta(u, \bullet)} X_t^{S}(\differential{u}) + \measureIntegral{X_t^{I}}{\gamma} \right) $. ii) Determine the event type by drawing a categorical random variable with probabilities equal to the ratios of the hazards of the individual events and the total hazard. A pseudocode for simulating a similar age-structured birth-death-transformation system is given in \cite{KhudaBukhsh2020Delay}. 
     
     In addition to the assumption of the global jump rates (in terms of the instantaneous intensity functions $\beta$ and $\gamma$) of the Markov process $X_t$ being bounded above by a positive finite quantity, we also assume the intensity functions $\beta$ and $\gamma$ are continuous. In order to study the \ac{FLLN} approximation of the scaled process $n^{-1}X_t$, we further assume a finite second moment condition on the initial population size. That is, we assume $\sup_{n} \Eof{n^{-2} \left( N_S(0) + N_I(0)  \right)^2} < \infty$. Finally, we assume the initial age distribution does not explode. 
     
     Note that the assumptions about the initial size of the population are satisfied because $n$ is chosen to be the size of the initial susceptible population and $m/n \to \rho \in (0, 1)$ as mentioned in \Cref{sec:model}.  With the above technical assumptions, we are now ready to study the moments of the stochastic process $X_t$ and associated martingale processes.

    \subsection{Moments and martingale properties}
    Note that the components $X_t^S, X_t^I$, and $X_t^R$ of $X_t$ satisfy the stochastic integral equations described in \eqref{eq:trajectory_equation}. Then, for a sufficiently large class of test functions $f: (a, s) \rightarrow f_s(a) $, the component measure-valued processes satisfy 
    \begin{align}
        \begin{aligned} 
        \measureIntegral{X_t^{S}}{f_t} &{} = \sum_{k=1}^{ N_{S}(0) } f_{t}({ t + \sigma_{k}(X_{0}^{S})  }) \\
        &{} \quad 
        - \int_0^t  \int_{\setOfNaturals} \int_0^\infty f_{t}({t-s + \sigma_{i}(X_{s-}^{S}) }) \indicator{i \le N_{S}(s-)} \indicator{\theta \le  \measureIntegral{X_{s-}^{I}}{ n^{-1} \beta(\sigma_{i}(X_{s-}^{S}), \bullet) }  } Q_1(\differential{s}, \differential{i}, \differential{\theta} ) \eqcomma \\
         \measureIntegral{X_t^{I}}{f_t} &{} = \sum_{k=1}^{ N_{I}(0) } f_{t}({ t + \sigma_{i}(X_{0}^{I})  }) \\
         &{} \quad 
         + \int_0^t  \int_{\setOfNaturals} \int_0^\infty f_{t}({t-s  }) \indicator{i \le N_{S}(s-)} \indicator{\theta \le \measureIntegral{X_{s-}^{I}}{ n^{-1} \beta(\sigma_{i}(X_{s-}^{S}), \bullet) }  } Q_1(\differential{s}, \differential{i}, \differential{\theta} ) \\
         &{} \quad - \int_0^t \int_{\setOfNaturals} \int_0^\infty f_{t}({t-s + \sigma_{k}(X_{s-}^{I}) }) \indicator{i \le N_{I}(s-)} \indicator{\theta \le \gamma( \sigma_{i}(X_{s-}^{I})) } Q_2(\differential{s}, \differential{i}, \differential{\theta}) \eqcomma \\
         \measureIntegral{X_t^{R}}{f_t} &{}=  \int_0^t \int_{\setOfNaturals} \int_0^\infty f_{t}({t-s }) \indicator{i \le N_{I}(s-)} \indicator{\theta \le \gamma ( \sigma_{i}(X_{s-}^{I})) } Q_2(\differential{s}, \differential{i}, \differential{\theta}) \eqcomma
        \end{aligned}
        \label{eq:moment_equation}
    \end{align}
    For different choices of the test function $f$, \eqref{eq:moment_equation} can be used to study various moments of the component measure-valued processes  $X_t^S, X_t^I$, and $X_t^R$. Moreover, \eqref{eq:moment_equation} allows us to study certain martingale processes associated with the stochastic process $X_t$. For susceptible, infected and recovered compartments, define the stochastic processes
    \begin{align}
        \begin{aligned}
            M_t^{S,f} & = \measureIntegral{X_t^S}{f_t} - \measureIntegral{X_0^S}{f_0} - \int_0^t \int_0^\infty \left( \partialDerivative{a}{f_s(a)}  + \partialDerivative{s}{f_s(a)} - f_s(a) \measureIntegral{X_s^I}{ \beta(a, \bullet) }  \right) X_s^S(\differential{a}) \differential{s} \eqcomma \\
            M_t^{I,f} & = \measureIntegral{X_t^I}{f_t}- \measureIntegral{X_0^I}{f_0} - \int_0^t \int_0^\infty \left( \partialDerivative{a}{f_s(a)}  + \partialDerivative{s}{f_s(a)} + f_s(0) \measureIntegral{X_s^I}{ \beta(a, \bullet) }  \right) X_s^S(\differential{a}) \differential{s} \\
            & \quad - \int_0^t \int_0^\infty \left( \partialDerivative{a}{f_s(a)}  + \partialDerivative{s}{f_s(a)} - f_s(a) \gamma(a)   \right) X_s^I(\differential{a}) \differential{s} \eqcomma \\
            M_t^{R, f} & = \measureIntegral{X_t^R}{f_t} - \measureIntegral{X_0^R}{f_0} - \int_0^t \int_0^\infty \left( \partialDerivative{a}{f_s(a)}  + \partialDerivative{s}{f_s(a)} + f_s(0) \gamma(a)   \right) X_s^I(\differential{a}) \differential{s} \eqstop 
        \end{aligned}
        \label{eq:martingale_processes}
    \end{align}
    Using the compensated \acp{PPM} of the original \acp{PPM} $Q_1$ and $Q_2$, we can show that the stochastic processes $M_t^{S, f}, M_t^{I, f}$, and $M_t^{R, f}$ are zero mean, square integrable, \cadlag martingale processes with predictable quadratic variations of the order $n^{-1}$. Here, we have used the fact that 
    \begin{align*}
    f_t(a+t-s) & = f_s(a) + \int_s^t \left( \partialDerivative{u}{ f_{u}(a+u-s)} + \partialDerivative{a}{f_{u}(a+u-s) } \right) \differential{u}  \eqstop 
    \end{align*}
    
    The trajectory equation for the scaled process $n^{-1} X_t$ can be written in a straightforward fashion by dividing both sides of \eqref{eq:trajectory_equation}. We can then write down moment equations like \eqref{eq:moment_equation} for the scaled process $n^{-1} X_t$ and also define the corresponding scaled martingale processes. Since the global jump rates are assumed to be bounded above by a positive finite quantity, the predictable quadratic variation processes vanish in the limit of $n \to \infty$. Therefore, in the limit of $n \to \infty$, we expect the scaled martingale processes to vanish, which, in turn, implies the scaled process $n^{-1} X_t$ converges to a deterministic, continuous function $x_t \defeq \left(x_t^S, x_t^I, x_t^R \right) $. However, such a convergence can only be guaranteed along a subsequence. Moreover, we need to ensure the sequence of the scaled processes $n^{-1} X_t$ is tight.  
    
    \subsection{Tightness of the scaled process and uniqueness of limit points}
    The two main instruments here are the Roelly criterion \cite{Roelly1986criterion} and the Aldous--Rebolledo criterion \cite{Joffe1986Semimartingale}. As done in \cite{Fournier2004microscopic} or \cite[Proposition~3.1]{Tran2008limit}, we can establish the required tightness by verifying the Roelly criterion in the vague topology and the Aldous--Rebolledo criterion for the sequence of the scaled stochastic processes $n^{-1} X_t$. The limit points $x_t \defeq \left(x_t^S, x_t^I, x_t^R \right) $ of the scaled process $n^{-1} X_t$ can be identified by virtue of the martingale representation in \eqref{eq:martingale_processes}. Indeed, the functions $x_t^S, x_t^I$, and $ x_t^R$ satisfy 
    \begin{align}
        \begin{aligned}
            \measureIntegral{x_t^S}{f_t} & = \measureIntegral{x_0^S}{f_0} + \int_0^t \int_0^\infty \left( \partialDerivative{a}{f_s(a)}  + \partialDerivative{s}{f_s(a)} - f_s(a) \measureIntegral{x_s^I}{ \beta(a, \bullet) }  \right) x_s^S(\differential{a}) \differential{s} \eqcomma \\
            \measureIntegral{x_t^I}{f_t} & = \measureIntegral{x_0^I}{f_0} + \int_0^t \int_0^\infty \left( \partialDerivative{a}{f_s(a)}  + \partialDerivative{s}{f_s(a)} + f_s(0) \measureIntegral{x_s^I}{ \beta(a, \bullet) }  \right) x_s^S(\differential{a}) \differential{s} \\
            & \quad + \int_0^t \int_0^\infty \left( \partialDerivative{a}{f_s(a)}  + \partialDerivative{s}{f_s(a)} - f_s(a) \gamma(a)   \right) x_s^I(\differential{a}) \differential{s} \eqcomma \\
            \measureIntegral{x_t^R}{f_t} & = \measureIntegral{x_0^R}{f_0} + \int_0^t \int_0^\infty \left( \partialDerivative{a}{f_s(a)}  + \partialDerivative{s}{f_s(a)} + f_s(0) \gamma(a)   \right) x_s^I(\differential{a}) \differential{s} \eqcomma
        \end{aligned}
        \label{eq:limiting_measure}
    \end{align}
    for a sufficiently large class of test functions $f: (a, s) \rightarrow f_s(a) $. Given that the initial measures $x_0^S, x_0^I$, and $x_0^R$ admit densities with respect to the Lebesgue measure, it can be shown that the functions $x_t^S, x_t^I$, and $ x_t^R$ admit densities with respect to the Lebesgue measure throughout a finite time interval $[0, T]$ for some $T >0$. Denoting the densities of the functions $x_t^S, x_t^I$, and $ x_t^R$ by $y_S(t, \bullet), y_I(t, \bullet)$, and $y_R(t, \bullet)$ respectively, we can see that the densities $y_S, y_I$, and $y_R$ satisfy the system of \acp{PDE} described in \eqref{eq:limiting_pde_1}. 
    
    Now, since we have assumed the global jump rates are bounded above by a finite positive number, we can show the solutions remain bounded on finite time intervals. In order to prove the uniqueness of the solutions, we can show that the distance between two possible solutions must vanish by invoking the Gr\"onwall's lemma and by virtue of the fact that the solutions remain bounded on finite time intervals.

    \section{Numerical scheme to solve the mean-field \acfp{PDE}}
    \label{appendix:pde_scheme}
    In this section we describe the numerical schemes used to solve the \acp{PDE}~\Cref{eq:sellke_pde}. In numerical terms, \Cref{eq:sellke_pde} is, inside the domain, an advection equation with one spatial dimension, in which the characteristics move at velocity $U(x,t)=1$, and with a forcing term given by the right-hand side term $-\gamma(s,t)y(s,t)$. Such equations are well-known and can be solved with an explicit Semi-Lagrangian scheme~\cite{SemiLagrangianschemes,Staniforth1990}. The potential source of numerical instability comes from the non-linear non-local boundary condition~\Cref{eq:sellke_pdeboundary}. We have opted for a numerical scheme which combines the explicit Semi-Lagrangian approach inside the domain and an implicit method to treat the solution at the boundary. Note that at the boundary we need to compute a scalar quantity; therefore, implementing an implicit method does not have a noticeable impact on the run-time of the numerical scheme itself, while improving the stability of the solution.\\
    We define a mesh with spacing $\Delta X = 1/M$ and $\Delta T = 1/N$, and points $s_i = i \Delta x $ and $t^n = n \Delta T$, with $0\leq i \leq M$ and $0\leq n \leq N$. The discretised set of equations is then:
    \begin{eqnarray}
    \frac{y( t^{n+1},s_i) - y(t^n,s_i - \frac{\Delta X}{\Delta T})}{\Delta T} &=& -\gamma\left(s_i-\frac{\Delta X}{2\Delta T}\right) y\left(t^n,s_i-\frac{\Delta X}{\Delta T}\right),\\
    \frac{x_S(t^{n+1}) - x_S(t^n)}{\Delta T} &=& - x(t^{n+1}) \sum_{k=1}^{M}\beta y(t^{n+1},s_k) \Delta x,\\ 
    \end{eqnarray}
    and, for the boundary condition
    \begin{equation}
        y(t^{n+1},0) = x(t^{n+1})\sum_{k=1}^{M} \beta y(t^{n+1},k) \Delta x.
    \end{equation}
    For simplicity, we use $\Delta X = \Delta T$, so that $s_i - \frac{\Delta X}{\Delta T} = s_{i-1}$. In~\ref{algo:pde}, we outline our implementation of the code.  This returns $z_s(t)$ and $z_I(t)=\int_0^\infty y_I(t,s) \, ds$. It is straightforward to modify it to return $y_I(t,s)$. 
    
 \begin{algorithm}[H]
 \begin{algorithmic}[1]
 \Require{$\gamma(s)$, $\beta$, $\rho$, $f(s)$, number\_of\_points, $T_f$}\\
 time\_mesh = [$n*T_f/$number\_of\_points  for $n$ in range(0,number\_of\_points)]    \\
 space\_mesh = time\_mesh \Comment{declare space and time meshes}\\       
 dx = 1/number\_of\_points \Comment{space step $\Delta x$}\\
 dt = dx\\
 y = zeros[time\_mesh] \Comment{allocate memory for I(t)}\\
 Y = zeros[space\_mesh] \Comment{Allocate memory to hold y[t][s] at every time step}\\
 $x_S$ = zeros[time\_mesh] \Comment{allocate memory for S(t)}\\
 x\_s[0] = 1 \Comment{initial fraction of susceptible people}\\
 \For{s in space\_mesh}\\
    \qquad Y = $\rho$  f[s] \Comment{initial condition on y[0][s]}\\
\EndFor\\
\\
A = zeros(space\_mesh,space\_mesh)
\For{s in space\_mesh}\\
\qquad A[s][s-1] = 1/(1+$dx$*$\gamma$[s-$\frac{1}{2}$]) \Comment{first order approximation of the \ac{PDE} operator}
\EndFor
\For{t in time\_mesh}\\
    \qquad  Y = A*Y  \Comment{PDE propagation}\\
    \qquad intY = sum($\beta[s]$*Y*dx) \Comment{$\int_0^\infty \beta(s) y(t+1,s)\,ds$ }\\
    
    \qquad x[t+1] = x[t]/(1+dx*intY) \Comment{update $x(t)$}\\
    \qquad y[t+1] = sum(Y)*dx  \Comment{$I(t) = \int y(t,s) ds$}\\
    
    \qquad y[0] = x[t+1] * intY \Comment{update Y at boundary with implicit scheme}\\
\EndFor\\
 \Return{$y$ and $x$}
 \end{algorithmic}
 \caption{Pseudo code to solve the PDE}
 \label{algo:pde}
\end{algorithm}
    
    \section{Distribution of $R_0$ estimates for \ac{FMD} and COVID-19}
    Here we report the estimates for $R_0$ from the bootstrap analysis of the \ac{FMD} data and COVID-19 Delta wave in India (see \Cref{sec:numerical}). Results are based on $500$ bootstrap samples obtained from simulating infection/recovery times with parameters given by the \ac{MLE}. 
    \begin{figure}[h]
        \centering
        \includegraphics{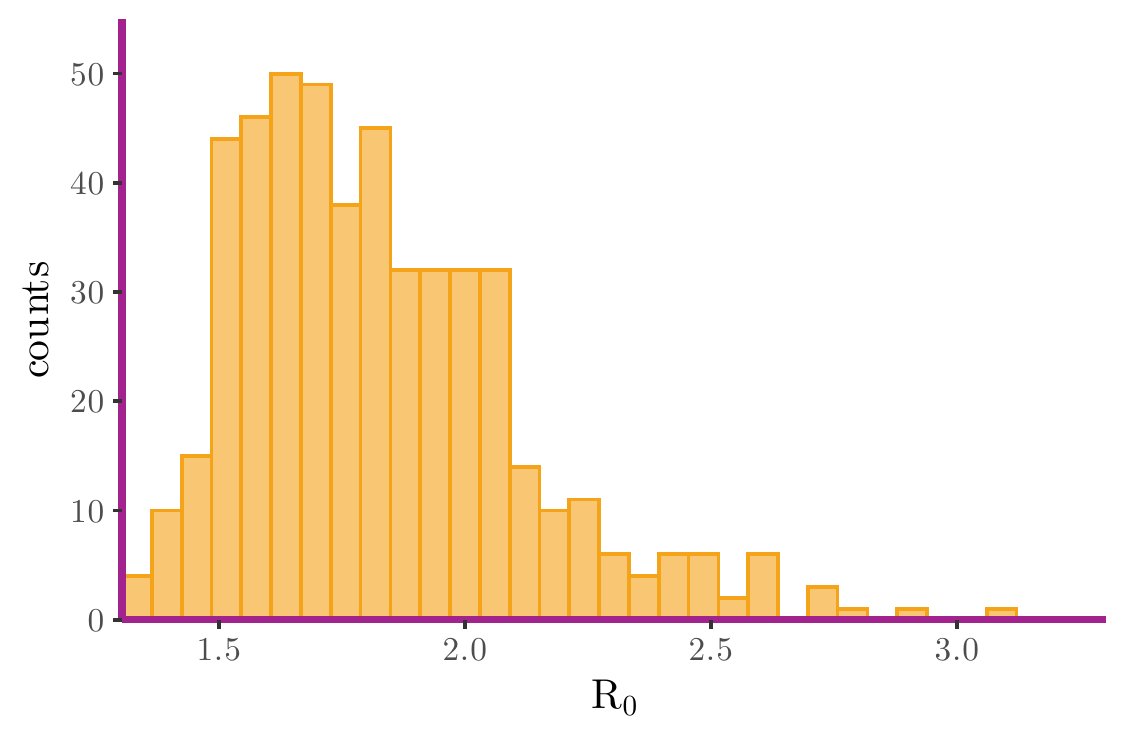}
        \caption{\ac{FMD} $R_0$ estimates from the bootstrap method.}
        \label{fig:R_0fmd}
    \end{figure}
    \begin{figure}[h]
        \centering
        \includegraphics{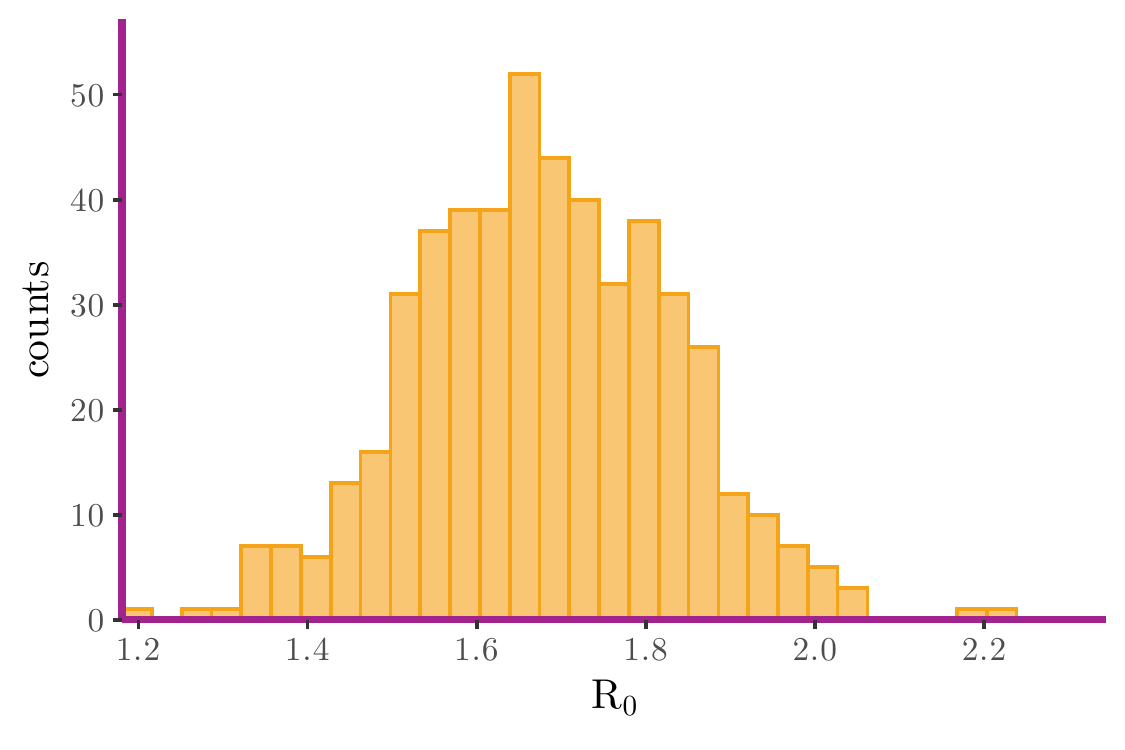}
        \caption{COVID-19 $R_0$ estimates from the bootstrap method.}
        \label{fig:R_0India}
    \end{figure}
    


    
    \section{Software}
    \label{appendix:software}
    A Python implementation of the methods is provided in \cite{GitHubLink} as a GitHub repository. 
    
    \section{Important notations}
    \label[Appendix]{appendix:tableofsymbols}
    
    \begin{table}[H]
      \centering
      \begin{tabular}{|| c | c ||}
        \hline 
        Notation & Meaning \\
        \hline 
        $\setOfNaturals$ & The set of natural numbers\\
        $\setOfReals$ & The set of reals \\
        $\setOfPositiveReals$  & The set of non-negative reals\\
        $\indicator{A}(x)$ & Indicator (characteristic) function of the set $A$\\
        $\delta_x$ & Dirac delta function at $x$\\
        $\borel{A}$ & The Borel $\sigma$-field of subsets of a set $A$\\
        $\mathcal{M}_{P}(E)$ & The space of finite point measures on the set $E$\\
        $D([0,T], E)$ & The space of  $E$-valued \cadlag functions defined on $[0,T]$\\
        $\measureIntegral{\mu}{f}$ & The integral $\int f \differential{\mu}$ \\
        \hline 
      \end{tabular}
    \end{table}

    \section{Acronyms}
    
\begin{acronym}[OWL-QN]
    \acro{ABC}{Approximate Bayesian Computation}
	\acro{ABM}{Agent-based Model}
	\acro{BA}{Barab\'asi-Albert}
	\acro{BD}{Birth-death}
	\acro{BM}{Brownian Motion}
	\acro{CCDF}{Complementary Cumulative Distribution Function}
	\acro{CDC}{Centers for Disease Control and Prevention}
	\acro{CDF}{Cumulative Distribution Function}
	\acro{CLT}{Central Limit Theorem}
	\acro{CM}{Configuration Model}
	\acro{CME}{Chemical Master Equation}
	\acro{CRN}{Chemical Reaction Network}
	\acro{CTMC}{Continuous Time Markov Chain}
	\acro{DTMC}{Discrete Time Markov Chain}
    \acro{DRC}{Democratic Republic of Congo}
	\acro{DSA}{Dynamic Survival Analysis}
	\acro{ER}{Erd\"{o}s-R\'{e}nyi}
	\acro{ESI}{Enzyme-Substrate-Inhibitor}
	\acro{FCLT}{Functional Central Limit Theorem}
	\acro{FIFO}{First In First Out}
	\acro{FJ}{Fork-Join}
	\acro{FLLN}{Functional Law of Large Numbers}
	\acrodefplural{FLLN}[FLLNs]{Functional Laws of Large Numbers}
	\acro{FMD}{Foot-and-Mouth Disease}
	\acro{FPT}{First Passage Time}
	\acro{GBP}{General Branching Process}
	\acro{HJB}{Hamilton–Jacobi–Bellman}
	\acro{iid}{independent and identically distributed}
	\acro{IPS}{Interacting Particle System}
	\acro{KL}{Kullback-Leibler}
	\acro{LDP}{Large Deviations Principle}
	\acro{LLN}{Law of Large Numbers}
	\acrodefplural{LLN}[LLNs]{Laws of Large Numbers}
	\acro{LNA}{Linear Noise Approximation}
	\acro{MABM}{Markovian Agent-based Model}
	\acro{MAPK}{Mitogen-activated Protein Kinase}
	\acro{MCMC}{Markov Chain Monte Carlo}
	\acro{MFPT}{Mean First Passage Time}
	\acro{MGF}{Moment Generating Function}
	\acro{MLE}{Maximum Likelihood Estimate}
	\acro{MM}{Michaelis--Menten}
	\acro{MPI}{Message Passing Interface}
	\acro{MSE}{Mean Squared Error}
	\acro{ODE}{Ordinary Differential Equation}
	\acro{PDE}{Partial Differential Equation}
	\acro{PDF}{Probability Density Function}
	\acro{PGF}{Probability Generating Function}
	\acro{PGM}{Probabilistic Graphical Model}
	\acro{PMF}{Probability Mass Function}
	\acro{PPM}{Poisson Point Measure}
	\acro{PRM}{Poisson Random Measure}
	\acro{psd}{positive semi-definite}
	\acro{PT}{Poisson-type}
	\acro{QSSA}{Quasi-Steady State Approximation}
    \acro{RBM}{Reflecting Brownian Motion}
	\acro{rQSSA}{reversible QSSA}
	\acro{s.d.}{Standard Deviation}
	\acro{SDS}{Survival Dynamical System}
	\acro{SEIR}{Susceptible-Exposed-Infected-Recovered}
	\acro{SI}{Susceptible-Infected}
	\acro{SIR}{Susceptible-Infected-Recovered}
	\acro{SIS}{Susceptible-Infected-Susceptible}
	\acro{SPDE}{Stochastic Partial Differential Equation}
	\acro{sQSSA}{standard QSSA}
	\acro{SSA}{Stochastic Simulation Algorithm}
	\acro{ssLNA}{Slow-scale Linear Noise Approximation}
	\acro{tQSSA}{total QSSA}
	\acro{WS}{Watts-Strogatz}
	\acro{whp}{with high probability}
\end{acronym}

    \section*{Acknowledgments}
    WKB was supported by the President's Postdoctoral Scholars Program (PPSP) of the Ohio State University.  EK and WKB were supported by the  National Institute of Allergy and Infectious Diseases (NIAID) Grant R01 AI116770, and GAR, EK and WBK were supported by the National Science Foundation (NSF) Grant DMS-2027001. WKB, EK and GAR also acknowledge the support of Mathematical Biosciences Institute (MBI) at the Ohio State University. IZK and FDL acknowledge support from the Leverhulme Trust for the Research Project Grant RPG-2017-370. IZK, FDL and MJ acknowledge the support of the Dr Perry James Browne Research Centre. The authors also wish to acknowledge Professor Michael Tildesley for providing \ac{FMD} daily incidence data. 
    
    \bibliographystyle{BibStyle_WKB}
    
    \bibliography{bibliography}
\end{document}